\documentclass{spie}%
\usepackage{array}
\usepackage{amsmath,graphicx,bbm}
\usepackage{mathrsfs}
\usepackage{subfigure}
\usepackage{ccaption}
\usepackage{color}
\usepackage{amssymb}
\usepackage{tikz}
\usepackage{float}
\usepackage{calc}
\usepackage{multirow}
\usepackage{natbib}

\newcommand{\procspie}{Proc. SPIE } 
 
\newcommand{\aap}{Astron. \& Astrophys. } 
\newcommand{\mnras}{M.N.R.A.S }

\newcommand{\Id}[1]{\mathbf{I}_{#1}}
\newcommand{\UN}[1]{[\mathbf{1}]_{#1}}
\newcommand{\ZERO}[1]{[\mathbf{0}]_{#1}}

\newcommand{\Conon}[0]{\mathcal{C}_\text{onon} }
\newcommand{\Conoff}[0]{\mathcal{C}_\text{onoff} }
\newcommand{\Coffoff}[0]{\mathcal{C}_\text{offoff} }

\newcommand{\cana}[0]{\textsc{Canary} }

\newcommand{\sig}[1]{\sigma_{\text{#1}}}
\newcommand{\sigdeux}[1]{\sigma^{2}_{\text{#1}}}
\newcommand{\cov}[2]{\mathcal{C}^{\text{#1}}_{#2} }
\newcommand{\tr}[1]{\text{tr}\para{#1}}
\newcommand{\module}[1]{\left\vert #1 \right\vert}
\newcommand{\para}[1]{\left(#1\right)}
\newcommand{\cro}[1]{\left[#1\right]}
\newcommand{\norme}[1]{\left\| #1 \right\|}
\newcommand{\aver}[1]{\left\langle #1 \right\rangle}

\newcommand{\xth}[1]{#1^{\text{th}}}
\newcommand{\hl}{h_{l}}
\newcommand{\sa}{\mathbf{S}_{\text{off}}}
\newcommand{\sbe}{\mathbf{S}_{\text{on}}}
\newcommand{\s}{\mathbf{S}}
\newcommand{\ns}{\text{n}_{\text{s}}}
\newcommand{\nwfs}{\text{n}_{\text{wfs}}}

\newcommand{\Nf}{\text{n}_{\text{f}}}

\newcommand{\nl}{\text{n}_{\text{l}}}
\newcommand{\nlgs}{\text{n}_{\text{lgs}}}
\newcommand{\nngs}{\text{n}_{\text{ngs}}}
\newcommand{\rz}{r_0}

\newcommand{\lz}{L_0}
\newcommand{\cnh}{C_n^2(h)}

\newcommand{\Mint}{\mathbf{M}_{\text{int}}}

\newcommand{\Msz}{\mathbf{M_{rz}}}

\newcommand{\R}{\mathbf{R}}
\newcommand{\Rglao}{\mathbf{R}^\text{Glao}}

\newcommand{\covnoise}{\mathcal{C}^{\text{Noise}}}
\newcommand{\covalias}{\mathcal{C}^{\text{Alias}}}
\newcommand{\tret}{\Delta_t}

\newcommand{\thcor}{\tilde{h}_{\text{cor}}}
\newcommand{\Dphi}[1]{D_{\phi}\left(#1\right)}
\newcommand{\kk}{\mathbf{k}}
\newcommand{\rhob}{\boldsymbol{\rho}}

\newcommand{\dipl}{\boldsymbol{\delta}_{ipl}}
\newcommand{\xipl}{x_{ipl}}
\newcommand{\yipl}{y_{ipl}}
\newcommand{\ri}{\mathbf{r}_{i}}
\newcommand{\alphap}{\boldsymbol{\alpha}_{p}}
\newcommand{\hlgsp}{h^{\text{LGS}}_{p}}

\newcommand{\djql}{\boldsymbol{\delta}_{jql}}
\newcommand{\xjql}{x_{jql}}
\newcommand{\yjql}{y_{jql}}
\newcommand{\Dijpql}{\boldsymbol{\Delta}_{ijpql}}
\newcommand{\gxx}{G_{xx}}
\newcommand{\gyy}{G_{yy}}
\newcommand{\gxy}{G_{xy}}

\newcolumntype{P}[1]{>{\centering\arraybackslash}p{#1}}

\title{Wave-front error breakdown in LGS MOAO \\ validated on-sky by CANARY}

\author[a]{O.A. Martin}
\author[b]{E. Gendron}
\author[b]{G. Rousset}
\author[b]{D. Gratadour}
\author[b]{F. Vidal}
\author[c]{T.J. Morris}
\author[c]{A.G. Basden}
\author[c]{R.M. Myers}
\author[a]{C.M. Correia}
\author[d]{D. Henry}

\affil[a]{Aix Marseille Universit\'e, CNRS, LAM, Laboratoire d'Astrophysique de Marseille, Marseille, France , 38 rue F. Joliot-Curie, 13388 Marseille Cedex 13, France}
\affil[b]{LESIA, Observatoire de Paris - Paris Sciences et Lettres, CNRS, Universit\'e Paris Diderot - Sorbonne Paris Cit\'e, Universit\'e P. et M. Curie - Sorbonne universit\'es, 5 pl. Janssen, 92190 Meudon,
	France}
\affil[c]{Centre for Advanced Instrumentation, Durham Univ., South Road, Durham, DH1 3LE, UK}
\affil[d]{UKATC, Royal Observatory Edinburgh, Blackford Hill, Edinburgh EH9 3HJ, UK}

\authorinfo{Further author information: olivier.martin@lam.fr}
\begin{document} 
	\maketitle
	\begin{abstract}
		
	\textsc{Canary} is the multi-object adaptive optics (MOAO) on-sky pathfinder developed in the perspective of Multi-Object Spectrograph on Extremely Large Telescopes~(ELTs). In 2013, \textsc{Canary} was operated on-sky at the William Herschel telescope~(WHT), using three off-axis natural guide stars~(NGS) and four off-axis Rayleigh laser guide stars~(LGS), in open-loop, with the on-axis compensated turbulence observed with a H-band imaging camera and a Truth wave-front sensor~(TS) for diagnostic purposes. Our purpose is to establish a reliable and accurate wave-front error breakdown for LGS MOAO. This will enable a comprehensive analysis of \cana on-sky results and provide tools for validating simulations of MOAO systems for ELTs. To evaluate the MOAO performance, we compared the \cana on-sky results running in MOAO, in Single Conjugated Adaptive Optics~(SCAO) and in Ground Layer Adaptive Optics~(GLAO) modes, over a large set of data acquired in 2013. We provide a statistical study of the seeing. We also evaluated the wave-front error breakdown from both analytic computations, one based on a MOAO system modelling and the other on the measurements from the \cana TS. We have focussed especially on the tomographic error and we detail its vertical error decomposition~(VED). We show that \cana obtained 30.1\%, 21.4\% and 17.1\% H-band Strehl ratios in SCAO, MOAO and GLAO respectively, for median seeing conditions with 0.66" of total seeing including 0.59" at the ground. Moreover, we get 99\% of correlation over 4,500 samples, for any AO modes, between two analytic computations of residual phase variance. Based on these variances, we obtain a reasonable Strehl-ratio~(SR) estimation when compared to the measured IR image SR. We evaluate the gain in compensation for the altitude turbulence brought by MOAO when compared to GLAO.
		
	\end{abstract}
	

\maketitle


\section{Introduction}
\label{S:Intro}

New scientific challenges will become achievable thanks to the future 30-40 m class telescopes such as the ESO European Extremely Large Telescope (E-ELT) \citep{Pherson2012}. In particular, it will be possible to track properties of first galaxies at high red-shifts \citep{Evans2008}. Such distant objects are smaller than the seeing disk, and so faint that they may require hours of integration time. They must be observable by integral field spectroscopy and statistics on these objects will be provided by a large multiplex system. Such a statistical analysis requires the observation of the greatest possible number of galaxies distributed over a very large field of view (FoV). Thanks to the large light-collecting power of the E-ELT combined with an adaptive optics (AO) system having high sky coverage, such observations will be feasible.

To perform the multi-object correction, an AO concept has been proposed \citep{Hammer2002}~: Multi-Object Adaptive Optics (MOAO). A MOAO system is composed of individual optical trains that split the FoV into different scientific directions. Each of these optical trains includes a single deformable mirror (DM) which corrects the turbulence over the small sub-field (of the order of an arc-second) encircling the galaxy. In addition in MOAO, objects to be observed are too faint to provide the number of photons required to measure the wave front (WF). The WF sensing is thus achieved thanks to other natural guide stars~(NGS) within the field and laser guide stars~(LGS). Unfortunately, the turbulence measured by WF sensors (WFS) decorrelates with angular distance because of the anisoplanatism effect \citep{Fried1982}. The mitigation of this effect requires a tomographic approach~: the DMs are driven from a combination of the off-axis WF measurements taking into account their angular correlation. Using several NGS and LGS in the FoV, MOAO estimates the wave-front in the scientific directions by a tomographic reconstruction to mitigate the anisoplanatism effect.

\textsc{Eagle}~\citep{Cuby2010} is a MOAO-assisted multi-object spectrograph~(MOS), that has been proposed for the E-ELT for the statistical study of the formation process of distant galaxies. The concept of \textsc{Eagle} is to cover a 7 arc-minutes diameter FoV and features 20 MOAO spectroscopic channels, working in parallel, and delivering a corrected wave-front for achieving an ensquared energy (EE) greater than 30~\% in a resolution element of 75 mas. The AO study \citep{Rousset2010} has demonstrated that the tomographic reconstruction must be based on a minimum of six LGS WFS and four to five NGS WFS in order to be able to reach the correction performance and the sky coverage requirement. Today, a new multi-object spectrograph, \textsc{Mosaic}, possibly equipped with MOAO, is entering in phase A study for the E-ELT~\citep{Hammer2014}.

To demonstrate the technical feasibility of MOAO sky observations, the pathfinder \cana~\citep{Myers2008} has been developed by an international consortium. It began in 2007, with first light in 2010 at the William Herschel Telescope~(WHT) on La Palma, in the Canary islands (Spain)~\citep{Gendron2011}. \cana is also a platform where innovative techniques and algorithms can be tested on-sky for AO control techniques \citep{Sivo2014,Osborn2014,Bitenc2015} and Shack-Hartmann~(SH) centroiding \citep{Basden2012,Basden2014} for instance. 

In this paper, we present the analysis of on-sky results acquired by \cana in 2013. We first describe the MOAO control and tomographic reconstruction with LGS in Sect.~\ref{S:phaseB}, as used with \cana. In particular, we discuss the implementation of the mixed NGS/LGS tomographic reconstruction.

In Sect.~\ref{S:tomoerror}, we introduce the tomographic error in LGS MOAO and describe how its evaluation is achieved for \cana. In Sect.~\ref{S3}, we present an analytic LGS MOAO error breakdown and describe how each term is derived. The variance of the residual phase taken in summing all terms, assumed to be independent, is compared to what we estimate from the truth sensor~(TS). We show in Sect.~\ref{S4} the two methods lead to the same evaluation of the residual phase, with a 99~\% correlation over a large set of 4,500 on-sky datasets. We also present an overview of the on-sky \cana performance in single conjugate AO (SCAO), MOAO, and open-loop ground layer AO (GLAO).

Finally in Sect.~\ref{S5}, we focus on particular results acquired by \cana running successively in SCAO, MOAO and GLAO modes. Using tools introduced in Sect.~\ref{S3}, we evaluate the \cana error breakdown in these modes and analyze the improvement achieved using LGS/NGS tomography.  We conclude in Sect.~\ref{S6}.

\section{\cana design~: phase B} 
\label{S:phaseB}
\subsection{Phase B design} \label{SS11}

The \cana pathfinder is deployed on one of the Nasmyth focii of the 4.2m WHT. The project has been planned with several phases of increasing system complexity \citep{Myers2008} to lead to a comprehensive demonstration of the MOAO configuration as foreseen for \textsc{Eagle} on the E-ELT~\citep{Rousset2010}. 

The \cana optical design during phase B \citep{Morris2013} includes up to eight SH WFS of 36 valid sub-apertures each ($7\times7$), three being off-axis NGS WFS, four being off-axis Rayleigh LGS WFS and one being an on-axis NGS WFS. They are installed behind an optical de-rotator (K mirror) placed at the entrance of the bench to compensate for the field rotation at the Nasmyth focus. The three off-axis NGS WFS are installed on a target acquisition system~(TAS) in the telescope focal plane which allows us to point at directions anywhere within a FoV of $2.5'$ diameter. The four Rayleigh LGS WFS have a selectable range gate (to tune the guide star height and depth), set to 21~km for the data presented in this paper, and are optically combined on a unique electronically time-gated CCD \citep{Morris2013,Morris2014}. The on-axis TS measures the WF corrected by the DM and tip-tilt~(TT) mirror for calibration procedures and for performance diagnostics. A dichroic plate splits the corrected WF beam into a visible part (400-900~nm) for the TS and an infra-red part (900-2500~nm) for an infra-red imaging camera, called \textsc{Camicaz}, equipped with a NICMOS detector and a filter wheel delivering J, H and K-band images \citep{Gratadour2013b}.

All the WFS cameras are triggered with a master clock derived from a sub-frequency of the laser pulses, leading to synchronous WFS measurements sampled at 150~Hz for all cameras. The real-time computer (RTC)~\textsc{DARC} \citep{Basden2010, Basden2016} acquires synchronously the WFS data and provides the user with time series of wave-front slopes and DM and TT commands. 

\subsection{Wave-front sensor~(WFS) slope vectors} \label{SS12}
\begin{table}[t]
	\caption{\textbf{Notation}}
	\centering
	\scriptsize
	\begin{tabular}{c p{6.cm}}
		\hline
		\textbf{Quantity} & \textbf{Signification} \rule[-2pt]{0pt}{10pt}\\ 
		\hline
		& \\
		$\aver{\mathbf{U}}$& Temporal average of vector $\mathbf{U}$ \\
		$\tr{U}$& Trace of the matrix $\mathbf{U}$\\
		$\Id{\ns}$ & $\ns \times \ns$ identity matrix\\
		$\UN{\ns}$ & $\ns \times \ns$ matrix full with 1.\\
		$\ZERO{\ns}$ & $\ns\times \ns$ matrix full with 0.\\
        $\norme{\mathbf{U}}$ & $L_2$ norm of vector $\mathbf{U}$\\
		$\R$ & Tomographic reconstructor \\
		$\sa$ & Vector concatenating slopes measured by off-axis WFSs \\
		$\sbe$ & Vector concatenating slopes measured by on-axis WFS \\
		$\Msz$ & Calibrated reconstruction matrix Slopes to Zernike\\
		$\ns$ & Number of measured slopes by a single WFS, $\ns=72$ in this paper \\
		$\Nf$ & Number of acquired frames\\
		$\nl$ & Number of discrete turbulent layers\\
		$\nwfs$ & Number of WFS, $\nwfs=7$ in this paper\\
		$\nngs$ & Number of NGS only WFS, $\nngs=3$ in this paper\\
		$\nlgs$ & Number of LGS only WFS, $\nlgs=4$ in this paper\\
		\hline
	\end{tabular}
	\label{T:1}
\end{table}
The RTC provides for each WFS a vector $\s_p$ of $\ns$~(see Table~\ref{T:1}) slopes~($\ns=2\times 36$) every WFS frame. The set of local WF slopes along x and y directions, measured by the $\xth{p}$ WFS with $\ns/2$ sub-apertures, is collected into a global slope vector $\s_p$~:
\begin{equation} \label{E:sp}
\s_p = \left(
\begin{aligned}
&S_{x_1}\\
& \hspace{.5cm} \vdots\\
&S_{x_{\ns/2}}\\
&S_{y_1}\\
& \hspace{.5cm} \vdots\\
&S_{y_{\ns/2}}\\
\end{aligned}
\right)
- \s^\text{ref}_p,
\end{equation}
where $ \s^\text{ref}_p$ is the vector of the reference slopes, previously calibrated in the laboratory \citep{Vidal2014}. 
To remove tip-tilt signal from LGS measurements, we subtract the average slopes  on both the x and y axis~:
\begin{equation} \label{E:slgsi}
\begin{aligned}
\s_\text{LGSp} = \left(\Id{\ns} -  \mathbf{F}\right)\times \s_p
\end{aligned}
\end{equation}
where $\Id{\ns}$ is the identity matrix of size $\ns$ and $\mathbf{F}$ is defined as~:
\begin{equation} \label{E:F}
\mathbf{F}=	\dfrac{2}{\ns}\begin{bmatrix} 
\UN{\ns/2} & \ZERO{\ns/2}\\
\ZERO{\ns/2} & \UN{\ns/2}\\
	\end{bmatrix},
\end{equation}
where $\UN{\ns/2}$ and $\ZERO{\ns/2}$ are squared matrices of size $\ns/2$ filled respectively with ones and zeros~(see Table~\ref{T:1}).

We denote $\sa$ as the array resulting from the concatenation of slope measurements in the off-axis directions. This therefore includes the slopes coming from 4 LGS and 3 NGS, and is of size $ (2\times36\times7) \times \Nf$. Likewise, we denote $\sbe$ as the array of size $\ns \times \Nf$ that concatenates the TS measurements on the on-axis source. We have~:
\begin{equation}
\sa = \left(
\begin{aligned}
&\s_\text{LGS1} (t=1) \cdots \s_\text{LGS1} (t=\Nf) \\
&\s_\text{LGS2}(t=1) \cdots \s_\text{LGS2} (t=\Nf)\\
&\s_\text{LGS3}(t=1) \cdots \s_\text{LGS3} (t=\Nf)\\
&\s_\text{LGS4}(t=1) \cdots \s_\text{LGS4} (t=\Nf)\\
&\s_\text{NGS5}(t=1) \cdots \s_\text{NGS5} (t=\Nf)\\
&\s_\text{NGS6}(t=1) \cdots \s_\text{NGS6} (t=\Nf)\\
&\s_\text{NGS7}(t=1) \cdots \s_\text{NGS7} (t=\Nf)\\
\end{aligned}
\right)
\end{equation}
\begin{equation}
\sbe = \left(
\s_\text{TS} (t=1) \cdots \s_\text{TS} (t=\Nf)\\
\right), 
\end{equation}
where the $\s_\text{NGSp}(t)$ and $\s_\text{TS}(t)$ are acquired according to Eq.~\ref{E:sp}, while $\s_\text{LGSp}$ comes from Eq.~\ref{E:slgsi}. In all the following, $\sa$ and $\sbe$ will be WFS slopes measured with a disengaged loop, which are measurements of uncompensated phase. 

We define $\s^\text{Eng}_\text{on}$ as the TS measurements acquired when loop is engaged. This represents a measurement of the residual phase. Each time series of $\s^\text{Eng}_\text{on}(t)$ gathers TS slopes following Eq.~\ref{E:sp}.

\subsection{NGS/LGS-based reconstructors}

\subsubsection{Ground layer compensation~(GLAO) reconstructor}

\cana has a ground layer compensation mode~(GLAO mode) with open-loop correction. It is comparable to the MOAO mode, except the reconstructor is simplified. The GLAO reconstructor averages the WFS signals, ignoring the tip and tilt from the lasers. It is computed as a sum of a tip-tilt~(TT) reconstructor $\Rglao_\text{TT}$ that averages the slopes on NGS WFS only, and a high order reconstructor $\Rglao_\text{HO}$ that filters out the tip-tilt on all measurements and averages the slopes. Mathematically we have~:
\begin{equation} \label{E:Rglao}
	\Rglao = \Rglao_\text{HO} + \Rglao_\text{TT},
\end{equation}
where $\Rglao_\text{TT}$ is the $\ns \times (\ns\times\nwfs)$~:
\begin{equation} \label{E:RglaoTTR}
\Rglao_\text{HO} = \dfrac{1}{\nwfs}
\underbrace{\left[\cro{\Id{\ns}-\mathbf{F}}
\hspace{.1cm} \cdots
\cro{\Id{\ns}-\mathbf{F}}\right]}_{\nwfs}
,
\end{equation}
and $\Rglao_\text{TT}$ the matrix averaging the slopes on NGS WFS only is designated as~:
\begin{equation} \label{E:RglaoTT}
\Rglao_\text{TT} = \dfrac{1}{\nngs}\left[
\underbrace{\ZERO{\ns} \hspace{.1cm} \cdots \ZERO{\ns}}_{\nlgs}, 
\underbrace{\mathbf{F}_{} \hspace{.1cm} \cdots \mathbf{F}_{}}_{\nngs}
\right]
\end{equation}
Calibration of references slopes, DM offsets, interaction and command matrices are described by \citep{Vidal2014}.

\subsubsection{Minimum mean square error~(MMSE) reconstructor} \label{SS21}
The on-axis phase is determined by minimizing the distance between the on-axis measurements and the linear combination of off-axis measurements propagated through the reconstructor $\R$~:
\begin{equation} \label{E1}
\varepsilon =  \aver{\norme{\sbe(t) - \aver{\sbe} - \R(\sa(t)-\aver{\sa})}^2},
\end{equation}
where $\aver{\mathbf{...}}$ denotes the temporal average over a time series. Ensuring $\partial \varepsilon/\partial\mathbf{R} = 0$ with $\sa$ and $\sbe$ assumed to be Gaussian stochastic processes,  we get~:
\begin{equation} \label{E2}
\R_\text{MMSE} = \underset{\mathbf{R}}{\text{argmin}}(\varepsilon) = \Conoff \Coffoff^{\dagger}
\end{equation}
with $\Conoff$ the $\ns \times (\ns\times\nwfs)$ spatial cross-covariance matrix between on/off measurements, while $\Coffoff$ denotes the $(\ns\times\nwfs) \times (\ns\times\nwfs)$ off-axis WFS spatial covariance matrix. They are defined by~:
\begin{equation} \label{E:covmat}
\begin{aligned}
&	\Conoff = \aver{(\sbe-\aver{\sbe})(\sa-\aver{\sa})^t}\\
&	\Coffoff = \aver{(\sa-\aver{\sa})(\sa-\aver{\sa})^t}.
\end{aligned}
\end{equation}
We note that the matrix $\Coffoff^{\dagger}$ in Eq.~\ref{E2} is the generalized inverse of the covariance matrix $\Coffoff$.

\subsubsection{Filtering of pupil average slope}

Laser guide stars~(LGS) WFS do not properly measure the pupil average slope (the overall angle of arrival), and we have explained that we subtract this from their measurements, as stated by Eq.~\ref{E:slgsi}.
The same filtering operation on the lasers needs to be taken into account in the tomographic reconstructor in order to get rid of undetermined modes.
We use a global approach (as opposed to techniques known as split tomography \citep{GillesJOSA2008} where the reconstruction is performed independently for LGS and NGS).

Before applying Eq.~\ref{E2}, we  rework the covariance matrix $\Coffoff$ and apply on each of its blocks $[\Coffoff ]_{pp'}$ between two wave-front sensors $p$ and $p'$ the following filtering operation:
\begin{equation}
	A^t [\Coffoff ]_{pp'} \, A'
\end{equation}
where $A = \Id{\ns}$ when $p$ is a NGS wave-front sensor, and $A = \Id{\ns} - \mathbf{F}$ in the case of a LGS wave-front sensor (and similarly for $A'$ and $p'$).
After filtering, the covariance matrix $\Coffoff$ should be understood as the covariance of measurements excluding the undetermined modes.

As a comparison with other studies addressing the same problem, we find it interesting here to point out the method used by \citep{TallonFrimEELTSpie2010}, who modifies the diagonal wave-front sensor blocks of the covariance matrix using
\begin{equation}
	[\Coffoff ]_{pp} + B
\end{equation}
with B the covariance matrix of undetermined modes measurements for a given LGS wave-front sensor $p$. This is equivalent to adding noise to the measured undetermined modes, which will automatically result in their filtering by the inversion process.

The inversion process is interesting and can be compared between the two methods. In the case of Tallon, the addition does not modify the rank of $\Coffoff$, while in our case filtering two degrees of freedom (tip and tilt) for each of our four LGS WFS creates eight null eigenvalues (associated to eight eigenmodes that span the subspace of our filtered modes). Therefore our matrix cannot be inverted directly. Instead, we perform a singular value decomposition (SVD) (which reduces to a diagonalization in this particular case of a square, symmetric, positive definite matrix), invert all non-zero eigenvalues, and set the inverse of null ones to zero. This is how the average slope of LGS wave-front gets filtered out in our case. We believe this approach is equivalent to Tallon's approach with the coefficients of matrix $B$ tending towards infinity.

We have chosen this method because the coefficients on B would have been difficult to adjust experimentally. The positioning of our LGS suffers from drifts, possibly from rotations of the asterism, that translate into long-term drifts of the average slopes that are quite empirical to turn into a covariance noise. For this reason, we prefer to completely eliminate of those modes.

In addition, we should mention that, in any case, the large dispersion of the eigenvalues of $\Coffoff$ (even before filtering of tilt) makes a direct inversion numerically risky, and that is it preferable to filter out the weakest eigenmodes, instead of attempting to take the inverse of their eigenvalues. This still requires an SVD.

\subsection{Reconstructor identification}
\label{Ident}
The matrix $\Conoff$ can not be determined on-sky when observing because the lack of photons forbids the use a TS in the scientific directions. In addition, empirical covariance matrices suffer from statistical convergence errors \citep{Martin2012} that make the tomographic reconstructor too specific to the particular time series from which it has been computed. 

For these reasons, we fit the on-sky measured $\Coffoff$ to a model of this matrix detailed in appendix.~\ref{A:1}. It provides analytic values of $\Conoff$ and $\Coffoff$ from a list of parameters~:
\begin{itemize}
	\item[$\bullet$] $\nl$~: number of discrete layers, set {ad hoc}.
	\item[$\bullet$] $\hl$~: altitude of layers
	\item[$\bullet$] $\rz(\hl)$~: strength of layers
	\item[$\bullet$] $\lz(\hl)$~: outer scale of layers
	\item[$\bullet$] Telescope tracking error and vibration~: adding of an isoplanatic tip-tilt over all NGS WFS
	\item[$\bullet$] WFS angular positions~: they are sufficiently well known using encoders information from the target acquisition system
	\item[$\bullet$] System mis-registration~(pupil shifts, rotation and magnification)~: well calibrated on bench
	\item[$\bullet$] Centroid gain variations
    \item[$\bullet$] Noise slope variance for each WFS, separately identified on the $\sa$ time series by computing its temporal autocorrelation. 
\end{itemize}
In the previous list, only the altitude, strength and outer scale of layers are retrieved, along with the tracking error. Other parameters are either set {ad hoc} or assumed to be known well enough. 

The retrieval of these parameters is ensured thanks to the Learn\&Apply algorithm~\citep{Vidal2010}. It is a least-square minimization of the distance between the empiric covariance and our model of that. The algorithm operates iteratively using a Levenberg-Marquardt method. For determining the turbulent profile on an engaged set of data, we estimate the disengaged TS measurements, $\sbe$ from the measured ones $\s^\text{Eng}_\text{on}$~(see appendix~\ref{B:1}). We thus have an additional WFS, centred on-axis, that allows us to increase the maximum field of view covered by tomography.

Previously we have used a model \citep{Vidal2010,Vidal2014} based on a Fourier transformation of the Von-K\'arm\'an spectrum to get the bi-dimensional (2D) spatial covariance of the wave-front slopes. This algorithm was particularly efficient when coupled to wave-front sensors with fixed sub-aperture sizes, because the computation using fast Fourier transforms (FFT) could be used to extract all covariance values from the same 2D map between two wave-front sensors with arbitrary separation, without requiring any interpolation of the map and provided the sub-aperture size being a multiple of the FFT sampling. 

However, the cone effect inherent to LGS induces layer-projected sub-aperture sizes and pupil sizes that vary with altitude, which breaks the efficiency of the implementation when it's a matter of computing covariances between LGS and NGS. An interpolation procedure could be implemented in order to access values that lay between covariance samples of the FFT, but unfortunately this kind of approximation does not go integrate with a fitting procedure that internally requires the computation of derivatives as finite differences of the covariance function. For this reason, we have developed a new analytic way to compute the covariances as it is presented in appendix~\ref{A:1}. This method allows us to get each coefficient of the matrix independently, which is a great advantage for faster computation for parallel implementation \citep{Gendron2014}.

\section{Evaluation of the tomographic error}
\label{S:tomoerror}
\subsection{Raw tomographic error}
\label{SS:rawtomo}

Tomographic reconstruction introduces a {tomographic error} when compared to single conjugate AO, to be included in the MOAO residual phase variance. In addition, the calibrated reconstructor, applied on a given data set on-sky, is usually different from the MMSE one given by Eq.~\ref{E2}, because the turbulence may evolve between the calibration time and the acquisition time. Moreover, the identification of the model parameters assumes the turbulence is composed by a few discrete layers, which introduces additional errors.

We define $\mathbf{e}(t)$ as the error between the on-axis measurement and its tomographic reconstruction from the off-axis measurements~:
\begin{equation}
\label{e}
	\mathbf{e}(t) = \sbe(t) - \R\sa(t),
\end{equation}
with $\R$ the on-sky tomographic reconstructor. The tomographic error can be derived by computing the trace of the error covariance matrix $C_{ee}$~:
\begin{equation}
\label{E:Cee}
C_{ee} = \aver{(\mathbf{e}(t) - \aver{\mathbf{e}})(\mathbf{e}(t)-\aver{\mathbf{e}})^t}.
\end{equation}
By replacing $\mathbf{e}$ in Eq.~\ref{E:Cee}  by its expression given in Eq.~\ref{e}, the covariance matrix $C_{ee}$ can be split into four terms as follows~:
\begin{equation}
\label{E:Cee2}
C_{ee} = \Conon - \Conoff\R^t - \R \Conoff^t + \R \Coffoff \R^t
\end{equation}
where $\Conon$ is the $\ns\times\ns$ spatial covariance matrix for the on-axis direction~:
\begin{equation} \label{E:conon}
	\Conon = \aver{(\sbe(t) - \aver{\sbe}) (\sbe - \aver{\sbe})^{t}}.
\end{equation}
In Eq.~\ref{E:covmat}, we define $\Conoff$ and $\Coffoff$. We see in Eq.~\ref{E:Cee2} that the reduction of the tomographic error is given by the two terms involving the covariance matrix $\Conoff$ between the on-axis and the off-axis measurements. If the correlation is very low between these measurements, in the case of strong anisoplanatism, the on-axis error is not reduced by these two terms but instead increased by the term $\R \Coffoff \R^t$: the tomographic reconstruction should be avoided. This may appear when considering the highest altitude turbulence layers. On the contrary, for low altitude layers, WFS measurements are strongly correlated and we are able to significantly reduce the on-axis WF variance by the tomographic reconstruction. 

For the purpose of the WF error breakdown analysis in this paper, we will reconstruct the slopes on the Zernike basis. For \cana, we consider a reconstruction of the polynomials 2 to 36 \citep{Vidal2014}. We denote $\Msz$ the corresponding reconstruction matrix. The WF error is then defined by the quadratic sum of the Zernike coefficients. 
We get the WF error $\sigdeux{}(C_{ee})$, denoted as the {raw tomographic error}, from the trace of the Zernike coefficient covariance matrix given by~:
\begin{equation}
\label{E:sigmatomoraw}
\sigdeux{}(C_{ee}) = \tr{\Msz C_{ee} \Msz^t}
\end{equation}

\subsection{Separating the aliasing error from the tomographic one}
\label{SS:aliastomo}

In Eq.~\ref{E:Cee2}, we compute the raw tomographic error covariance matrix which may include a number of error terms like noise, aliasing and tomography. In this section, we will focus on the two spatial error terms related to the turbulence: the first linked to the low spatial frequencies of the turbulence well measured by the WFS and well compensated by the DM and the second linked to the high frequencies aliased in the measurements because of the spatial sampling by the WFS. A detailed analysis of aliasing in tomography can be found in~\cite{Quiros2010}.
To build the error breakdown, we want to separate the aliasing error from the tomographic one. For that purpose, we compute the aliasing error covariance matrix as~: 
\begin{equation}
\label{E:C-alias}
\covalias_{ee} = \Conon^\text{Alias} - \Conoff^\text{Alias}\R^t - \R (\Conoff^\text{Alias})^t + \R \Coffoff^\text{Alias} \R^t
\end{equation}
where the covariance matrices, $\Conon^\text{Alias}$, $\Conoff^\text{Alias}$ and $\Coffoff^\text{Alias}$, are derived by filtering in the turbulence spectrum all the low spatial frequencies below the DM cut-off frequency $1/2d$ with $d$ the DM actuators pitch, as described in Appendix~\ref{A:2}. In Eq.~\ref{E:C-alias}, the first term is related to the WFS placed on axis, the TS for \cana. The fourth term is the error due to the aliasing of the off-axis WFS propagated through the tomographic reconstruction. The two other terms are the correlation between the aliasing in the TS and in the off-axis WFS. This correlation is only significant for the ground layer where the same turbulence is sampled and measured by all the WFS. Hence these two terms reduce the aliasing error in the measurement of the TS. We will make use of Eq.~\ref{E:C-alias} to derive the aliasing contributors in the error breakdown established in the next Section. 

We now derive the tomographic error subtracted by the aliasing error. We define the tomographic error covariance matrix expressed for the low spatial frequencies of the turbulence (the WF low orders controlled by the DM) as:
\begin{equation}
C^\parallel_{ee} = C_{ee} - \covalias_{ee}
\end{equation}
And using Eqs.~\ref{E:Cee2} and~\ref{E:C-alias}, we find: 
\begin{equation}
\label{E:C-parallel}
C^\parallel_{ee} = \Conon^\parallel - \Conoff^\parallel\R^t - \R (\Conoff^\parallel)^t + \R \Coffoff^\parallel \R^t
\end{equation}
All these new covariance matrices $C^\parallel$ are deduced from the subtraction of the slope covariance matrices, defined in Section~\ref{SS:rawtomo} and computed as presented in Appendix~\ref{A:1}, by the aliasing covariance matrices $\covalias$ defined above. Hence $C^\parallel_{ee}$ can be fully computed from the identified model of atmosphere, as presented in Section~\ref{Ident}, here excluding the noise contributions. Using Eq.~\ref{E:sigmatomoraw}, we find the tomographic error of interest~: 
\begin{equation} 
\label{E:TomoIR}
\sigdeux{Tomography} = \sigdeux{}(C^\parallel_{ee}).
\end{equation}

To further describe the tomographic error decomposition, \citep{Gendron2014} has introduced the VED - for Vertical Error Distribution - as the decomposition of the WF tomographic error with altitude. Thanks to the statistical independence of the turbulent layers, the total tomographic error is the sum of the tomographic error on each turbulent layer at altitude $h_l$~:
\begin{equation} \label{E:cee3}
\begin{aligned}
\sigdeux{}(C^\parallel_{ee}) & = \sum_{l=1}^{n_l}\tr{\Msz C^\parallel_{ee}(\hl) \Msz^t}\\
& =\sum_{l=1}^{n_l} \sigdeux{}(C^\parallel_{ee}(\hl)).
\end{aligned}
\end{equation}
Such a decomposition allows us to identify which altitude layer contributes mostly to the tomographic error, and gives us the ability to quantify the robustness of reconstructor to profile variability \citep{Gendron2014}.

\section{Error breakdown in MOAO}
\label{S3}

The design of future AO-assisted MOS for ELTs will require detailed numerical simulations of AO systems with a very large number of degrees of freedom. To prepare those simulations, it is necessary to validate the current modelling of the MOAO systems and therefore to build an accurate WF error breakdown. \cana provides two sources of performance diagnostics : the residual phase as measured by the TS and the sky point spread function (PSF) as delivered by the imaging camera \textsc{Camicaz}. 

We propose two methods to evaluate the WF variance $\sigdeux{$\varepsilon$}$ as observed by the science camera. The first one as the independent terms~(IT) method gives $\sigdeux{$\varepsilon$}$ from \cana telemetry data by splitting the residual phase into a sum of terms assumed to be uncorrelated. We compare the IT approach with the TS method, as the direct evaluation of $\sigdeux{$\varepsilon$}$ based on the TS measurements. Our goal is then to compare the two methods on a large number of data sets acquired on-sky by \cana and to demonstrate that each of them leads to a very close evaluation of $\sigdeux{$\varepsilon$}$. 

\subsection{Analytic method~: the IT method}
\label{SS32}

The IT method consists of splitting the residual phase variance $\sigdeux{$\varepsilon_\text{IT}$}$ in a sum of error terms assumed to be statistically independent and coming from the AO system performance known limitations~:
\begin{equation}\label{E:DTI}
\sigdeux{$\varepsilon_\text{IT}$} =\sigdeux{Tomography} + \sigdeux{Aliasing} +\sigdeux{Noise} + \sigdeux{Servo}
+  \sigdeux{Fitting}+ \sigdeux{Go-to} + \sigdeux{Static} + \sigdeux{NCPA},
\end{equation}
where~:
\begin{itemize}
	\item[$\bullet$] $\sigdeux{Tomography}$ is the tomographic error on the WF low orders compensated by the system
	\item[$\bullet$] $\sigdeux{Aliasing}$ is the aliasing error from the off-axis WFS propagated through $\R$
	\item[$\bullet$] $\sigdeux{Noise}$ is the noise error from the off-axis WFS propagated through $\R$ and the MOAO loop
	\item[$\bullet$] $\sigdeux{Servo}$ is the temporal error due to the system transfer function including all delay
	\item[$\bullet$] $\sigdeux{Fitting}$ is the uncorrected high order error
	\item[$\bullet$] $\sigdeux{Go-to}$ is the go-to error of the DM
	\item[$\bullet$] $\sigdeux{Static}$ is the on-axis uncontrolled and uncalibrated static, or quasi-static error
	\item[$\bullet$] $\sigdeux{NCPA}$ is the non common path aberrations~(NCPA) residuals
\end{itemize}

Each of those variances involved in Eq.~\ref{E:DTI} are mostly computed using analytical equations, listed in the following subsections, depending on parameters deduced from the calibration data and the RTC telemetry data, which are $\sa(t)$ and $\sbe^\text{Eng}(t)$ when the loop is engaged.

\subsubsection{Tomographic error $\sigdeux{Tomography}$}
\label{SS:tomo}
The tomographic error is the error on the WF low orders compensated by the system, linked to the tomographic process, the used reconstructor $\R$ and the $\cnh$ profile encountered during the observation. This error $\sigdeux{Tomography}$ is given by Eq.~\ref{E:TomoIR}. 
The interest of our approach is to evaluate it without limiting statistic convergence effects specific to the particular acquired time series involved. Compared to \citep{Vidal2014}, the TS measurement is no longer required. It is only needed to compute the error covariance matrix for the low order modes, $C^\parallel_{ee}$, from the identified model of the atmosphere free from any noise and aliasing contribution~(see Sect.~\ref{SS:aliastomo}).

\subsubsection{Aliasing error $\sigdeux{Aliasing}$}
\label{SS:alias}
To evaluate the aliasing term, we only need to consider the off-axis aliasing error propagation through the reconstructor on the WF observed by the IR camera. From Eq.~\ref{E:C-alias}, we can extract this term, i.e. $\R\covalias_{\text{offoff}}\R^t$. The aliasing error variance is then given by~:
\begin{equation} \label{E:sigAliasIR}
 \sigdeux{Aliasing} = \tr{\Msz  \R \covalias_{\text{offoff}} \R^t \Msz^t}.
\end{equation}

The off-axis aliasing is also time-filtered by the system, which is filtering out the high temporal frequencies. It is reasonable, for a first approximation, to neglect the impact of the filtering on the aliasing error variance, which means to neglect the temporal high frequencies from the aliasing with regards to the others terms of the error breakdown.

\subsubsection{Noise error $\sigdeux{Noise}$}
\label{SS:noise}

As the aliasing, the noise in the residual WF error is due to the noise of the off-axis WFS propagated through the reconstructor $\R$ and the system. Considering the noise as a white process, the propagation through the loop can be readily taken into account. We denote $\sigdeux{Noise} $ as the variance noise contribution in the residual error variance. $\sigdeux{Noise} $ is derived by:
\begin{equation} \label{E:62}
\sigdeux{Noise} = \eta \times \tr{\Msz \R \covnoise_{\text{offoff}} \R^t \Msz^t},
\end{equation}
with~:
\begin{equation} \label{E:eta}
\eta = \dfrac{g}{2-g}(1-2g\tret(1-\tret)),
\end{equation}
where $g$ is the loop gain and $\Delta_t$ the system latency additional to the WFS exposure-time, assumed to be varying between 0 and 1 frame. From bench calibration, we get $\Delta_t = 0.45$ frame. 

The noise covariance matrix $\covnoise_{\text{offoff}}$ is assumed to be diagonal since the LGS spots elongation is negligible on \cana~\citep{Morris2013}. It is identified slope by slope computing the temporal autocorrelation on the slope time series. We extract the noise variance as a Dirac value at null delay through a parabolic fit of the turbulence contribution.

\subsubsection{Servo-lag error $\sigdeux{Servo}$}
\label{SS:servo}
\cite{Vidal2014} has proposed to base the calculation of the servo-lag error on a end-to-end approach, particularly to take into account a fractional system delay, but it does not include the propagation of the off-axis WFS measurements through the reconstructor $\R$.

We start by modelling the time-filtering process of the MOAO loop using the correction transfer function $\thcor(z)$ derived by a z-transform~(see appendix.~\ref{B:1})~:
\begin{equation} \label{E:thcormoao}
\thcor(z) = 1 - g\dfrac{\tret + (1-\tret)z}{z(z-1+g)}
\end{equation}
with $z = e^{2\pi i \nu/\nu_e}$ where $\nu_e$ is the time frequency sampling. The digital filter $\thcor(z)$ operates on the temporal spectrum $\R\mathbf{\tilde{\s}_\text{off}}$. We note $\mathbf{a}_\text{off}$ the atmospheric off-axis modes filtered by the tomographic reconstructor and the MOAO loop. Its temporal spectrum is derived by~:
\begin{equation} \label{E:sigdeuxBW}
\mathbf{\tilde{a}}_\text{off}(\nu) =  \thcor(z = e^{2i\pi\nu/\nu_e}) \times \Msz.\R.\mathbf{\tilde{\s}_\text{off}}(\nu),
\end{equation}
In Eq.~\ref{E:sigdeuxBW}, $\mathbf{\tilde{\s}_\text{off}}(\nu)$ includes a noise contribution from the off-axis WFS through the reconstructor and the correction transfer function $\thcor$. For de-noising the variance of $\mathbf{a}_\text{off}$,  
we subtract the variance of the off-axis noise, $\tr{\Msz \R \covnoise_{\text{offoff}} \R^t \Msz^t}$, multiplied by a coefficient which is the modulus of the MOAO correction transfer function, integrated along temporal frequencies. It has been computed by \cite{Vidal2014} to be $1+\eta$~(see. Eq.~\ref{E:eta}).

The servo-lag WF error is then derived by integrating the modal spectrum over temporal frequencies domain and modes, until the $\xth{36}$ one considered as the higher order to be correctable by the system. We have~:
\begin{equation} \label{E:servolag}
\sigdeux{Servo} = \norme{\int_{\nu}\mathbf{\tilde{a}}_\text{off}(\nu) d\nu}^2 - (1+\eta) \times \tr{\Msz \R \covnoise_{\text{offoff}} \R^t \Msz^t}.
\end{equation}

In Eq.~\ref{E:servolag} We do not extract the contribution of the off-axis aliasing. This term comes from the vector $\mathbf{\sa}$ and is propagated through the tomographic reconstructor and then through the MOAO correction transfer function. This later behaves as a high-pass filter. It means we neglect the contribution from the high temporal frequencies of the aliasing. For a Taylor-like turbulence, these frequencies are related to high spatial frequencies. Considering the turbulence spatial PSD is falling as $k^{-11/3}$, the high temporal frequency contribution of the aliasing we get in Eq.~\ref{E:servolag} is negligible when compared to the parallel modes.

\subsubsection{Fitting error $\sigdeux{Fitting}$}
\label{SS:fitting}

From end-to-end simulations, \citep{Vidal2014} has established for the DM of \cana the fitting error expression versus the DM actuators pitch $d$ and the Fried's parameter $r_0$~:
\begin{equation} \label{E:sigdeuxFit}
\sigdeux{Fitting} = 0.3125\para{\frac{d}{r_{0}}}^{5/3}
\end{equation}

\subsubsection{Go-to error $\sigdeux{Go-to}$}
\label{SS:goto}

Go-to errors are produced because of the open-loop control of DMs. Generally, the DM behaviour is modelled using an interaction matrix. Many ways exist to perform this calibration, and this has been investigated for the \cana case~(\cite{Kellerer2012}). In closed-loop AO systems, the DN mis-registration or inaccuracies in the interaction matrix calibration are compensated iteratively by the AO loop, thanks to the feedback between WFS and DM.  However, this is not the case for open-loop systems such as \cana.

We therefore have an additional WF error $\sigdeux{Go-to}$ that contributes to the residual phase variance. In \citep{Kellerer2012}, this go-to error is derived as a function of the commanded WF for several methods of computation of the interaction matrix. 

In the \cana case, this matrix has been computed using a sinusoidal method and, from results of \citep{Kellerer2012}, we propose to define the go-to error in taking 5~\% of the DM command. We have thus~:
\begin{equation} \label{E:sigdeuxOL}
\sigdeux{Go-to} = 0.05^2 \, \tr{\Msz \R \Coffoff\R^t\Msz^t}.
\end{equation}
This error is very small compared to the other error terms as we will see in Sect.~\ref{S5}.

\subsubsection{Static aberration errors $\sigdeux{Static}$}
\label{SS:static}

Differential quasi static aberrations may appear during an observation. A main reason is  the optical beams to the wave-front sensors and to the IR camera do not exactly cross the same areas of the optical surfaces of, both the telescope and the \cana bench. They include the field aberrations of the telescope and of the \cana de-rotator and the {creeping effect}. This later is a drifting of the DM shape with time~\cite{Kellerer2012}.

In order to mitigate the effects of static aberration error, we could use a model of the telescope field aberrations and de-rotator optical aberrations, as well as a {figure sensor} to monitor the DM shape ; however this approach was deemed too challenging to be applied in \cana. Instead, we perform on-sky a number of calibrations and related pre-compensations of these static aberrations for the different channels both before and during the observation. The procedure is given in~\citep{Vidal2014}.

Nevertheless in \cana, the calibration is not perfect and we are not able to update the static aberration compensation during the observation. Hence we have to include, in the WF error breakdown, a static aberration error term $\sigdeux{Static}$. We evaluate post-facto this error by computing~: 
\begin{equation} \label{E:64}
\sigdeux{Static} = \norme{\aver{\Msz \sbe^\text{Eng}(t)}_t}^2.
\end{equation}

\subsubsection{NCPA residual errors $\sigdeux{NCPA}$}
\label{SS:ncpa}
Non-common path aberrations (NCPA) are calibrated in closed loop on the TS and the bench artificial sources using a phase diversity algorithm processing the IR camera images \citep{Gratadour2013b}. The measured NCPA are then applied on the DM offset voltages during the whole duration of the observation as explained by \citep{Vidal2014}. The NCPA calibration errors are mainly dominated to the high spatial frequencies of the DM surface quality not correctable by the DM. They lead to a 80\% best Strehl ratio (SR) as measured in the 1550nm image at the end of the phase diversity procedure. This corresponds to an error of 115nm rms.

\subsection{Empirical method~: the TS method}
\label{SS33}
The empirical method consists in considering the TS as the primary source of information for determining the system performance. We denote $\sigdeux{TS}$ the variance of the TS WF engaged measurements that includes both static and dynamic terms~:
\begin{equation} \label{E:sigTS}
\sigdeux{TS} = \aver{\norme{\Msz\sbe^\text{Eng}(t)}^2}_t
\end{equation}

We expect the TS method to give an accurate performance estimation when compared to the science image, since only a dichroic plate separates the TS from the science path. The differences between the science wave-front and that measured by the TS are the TS noise, the TS aliasing, the fitting errors due to the wave-front high frequencies and NCPA residual errors. We want to estimate the variance of the science wave-front, and it can be written as~: 
\begin{equation} \label{E:DMTS}
\begin{aligned}
&\sigdeux{$\varepsilon_\text{TS}$} =\sigdeux{TS} - \sigdeux{NoiseTS} - \sigdeux{AliasingTS} \\
&  +\sigdeux{Fitting}+ \sigdeux{NCPA} 
\end{aligned}
\end{equation}
where $\sigdeux{NoiseTS}$ is the variance of the noise and $\sigdeux{AliasingTS}$ the variance of the aliasing, in the TS measurements. 

The parameter $\sigdeux{NoiseTS}$ is estimated using the noise covariance matrix $\covnoise_\text{onon}$, identified using the method based on the temporal autocorrelation of the time series of $\sbe^\text{Eng}$. Then, the TS noise variance term is given by~: 
\begin{equation} \label{E:signoiseTS}
\sigdeux{NoiseTS} = \tr{\Msz \covnoise_{\text{onon}}\Msz^t}
\end{equation}

The aliasing included into the TS measurements is firstly given by the conventional term linked to $\covalias_\text{onon}$ and is partially correlated to the off-axis aliasing propagated through the tomographic reconstructor. As presented in Section~\ref{SS:tomo}, the TS aliasing variance term is given by~: 
\begin{equation} \label{E:sigAliasTS}
\sigdeux{AliasingTS} =  \tr{\Msz (\covalias_{\text{onon}} - \covalias_{\text{onoff}}\R^t - \R (\covalias_{\text{onoff}})^{t}) \Msz^t}
\end{equation}
In this expression, the correlated part of the aliasing due to the ground layer is directly taken into account without any additional computation, compared to the previous approach given in \citep{Vidal2014}. 

\subsection{SCAO error breakdown for \cana}

Error breakdown in SCAO has been investigated for over 20 years~(see \cite{Rigaut1991, Gendron1994}). In SCAO, the TS is obviously available and it makes sense to estimate the residual phase variance using the TS measurements. We can directly use Eq.~\ref{E:DMTS}, but here not subtracting the term $\sigdeux{AliasingTS}$ which has to be kept in $\sigdeux{TS}$. 

In addition, we want to have the same breakdown as in Eq.~\ref{E:DTI}, but specific to SCAO. In SCAO, we will not have tomographic, static aberrations and go-to errors. The residual phase variance can be expressed as follows in SCAO mode~:
\begin{equation}\label{E:DTI_SCAO}
\sigdeux{$\varepsilon_\text{IT}$} = \sigdeux{AliasingTS} +\sigdeux{NoiseTS} + \sigdeux{Servo}+ \sigdeux{Fitting}+ \sigdeux{NCPA}.
\end{equation}
Values of the fitting error $\sigdeux{Fitting}$ and NCPA error $\sigdeux{NCPA}$ are discussed respectively in Sect.~\ref{SS:fitting} and~\ref{SS:ncpa}. 

We compute the aliasing error term $\sigdeux{AliasingTS}$ by projecting the covariance matrix of the aliasing in open-loop, derived from known turbulence parameters, onto the Zernike basis~:
\begin{equation} \label{E:aliasSCAO}
\sigdeux{AliasingTS} = \tr{\Msz \cov{Alias}{\text{onon}}\Msz^t}.
\end{equation}
In Eq.~\ref{E:aliasSCAO}, we neglect the propagation of the aliasing through the closed loop and the cross-correlation with parallel modes as well. But once again, {a posteriori} over the large amount of SCAO data we have processed, this approximation is not critical.

The noise error $\sigdeux{NoiseTS}$ is retrieved by the temporal auto-correlation function of the TS measurements. To make this estimation more accurate, we handle the estimation of the TS slopes in loop disengaged, $\sbe$ from the engaged ones $\sbe^\text{Eng}$~(see appendix~\ref{B:2}). The value of the noise we estimate is then multiplied by a propagation factor $1+g/(2-g)(1-2g\Delta_t(1-\Delta_t))$~(\cite{Vidal2014}) to take into account both the noise level at the sequence $t$, and the one from the sequence $t-1$ that has been propagating through the loop to be measured again by the TS.

Finally, the servo-lag error $\sigdeux{Servo}$ is derived identically to that in Eq.~\ref{E:servolag}~:
\begin{equation}
\sigdeux{Servo} = \int_{\nu}\norme{\mathbf{a}_\text{on}}^2(\nu) d\nu - \sigdeux{NoiseTS},
\end{equation}
where $\mathbf{\tilde{a}}_\text{on}$ is estimated from the TS measurements~:
\begin{equation} \label{E:aon}
\mathbf{\tilde{a}}_\text{on}(\nu) =  \tilde{h}_\text{cor}(\nu) \times \Msz\mathbf{\tilde{\s}_\text{on}}(\nu),
\end{equation}
where $ \tilde{h}_\text{cor}(\nu)$ is the correction transfer function given by \cite{Gendron1994} for SCAO systems.

\subsection{Evaluation of the Strehl ratio}
\label{SS34}
SR is known to be related to the phase variance by the Mar\'echal approximation~:
\begin{equation} \label{E:SRmar}
\widehat{\text{SR}}_\text{mar} = \exp(-(2\pi/\lambda)^2\sigdeux{$\varepsilon$}),
\end{equation}
where $\lambda$ is the wavelength of the science image and $\sigdeux{$\varepsilon$}$ can be either $\sigdeux{$\varepsilon_\text{IT}$}$ or $\sigdeux{$\varepsilon_\text{TS}$}$. Such an approximation is valid for small variance~$(\sigdeux{$\varepsilon$}<1 \text{rad}^2$), or high SR greater than 10-20\%. MOAO is not designed to get very high performance and the Mar\'echal approximation is not be accurate enough for our purposes. We propose to use the \citep{Parenti1994} heuristic formula~: 
\begin{equation}\label{E:SR}
\widehat{\text{SR}}_\text{par}  = \dfrac{{\mathrm e}^{- (2\pi/\lambda)^2\sigdeux{$\varepsilon_\text{HO}$} }}{1+ (2\pi/\lambda)^2\sigdeux{$\varepsilon_\text{TT}$}} + \dfrac{1-{\mathrm e}^{- (2\pi/\lambda)^2\sigdeux{$\varepsilon_\text{HO}$} }}{1+(D/r_0 (\lambda))^2}
\end{equation}
where $\sigdeux{$\varepsilon_\text{TT}$}$ and $\sigdeux{$\varepsilon_\text{HO}$}$ are the residual WF error on respectively the tip-tilt only and the higher orders~(tip-tilt removed).

Contrary to the Mar\'echal approach, the SR estimated from Eq.~\ref{E:SR} includes two terms. The first one is the contribution of the central core of the partially AO-corrected PSF but broadened by the residual TT variance, while the second one is the contribution of the PSF broad halo. The Parenti approximation, as we will see, gives a better estimation for SR lower than 30~\% than the Mar\'echal one , which is exactly the range of performance we have encountered with \cana.

\section{On-sky validation of MOAO}
\label{S4}
\subsection{Observation conditions}
\label{SS41}
We now focus on the \cana on-sky results acquired during the observing runs of phase B. Our purpose is firstly to prove the feasibility of managing a mixed LGS plus NGS MOAO system, secondly to evaluate the contribution of the tomographic reconstruction in the final performance of \cana. We have compared SR and error breakdown between several AO modes~: SCAO (closed-loop on TS), NGS MOAO, NGS plus LGS MOAO, TT plus LGS MOAO and GLAO (ground layer compensation only).

We have simultaneously recorded system telemetry (2048 frames at 150 Hz) and H-band images. Each science image is the sum of 15 individual exposures of one second, with no recentring in order to keep tip-tilt errors between individual exposures.\\

During the phase B runs, we have mostly observed three different NGS asterisms, that are presented in Table~\ref{T:ast}. In addition to the NGS, we also made use of four Rayleigh LGS. They are placed on a square asterism centred on-axis. Each LGS is located at 22.6" off-axis from the TS position. The distance of LGS, as defined by the range gate on the LGS WFS, is 21~km, regardless of the airmass, for all  nights. The gate width corresponds to a depth of 1.5~km in the atmosphere. The number of detected photons per LGS beacon is of the order of 300 photons by sub-aperture and by frame~\citep{Morris2013}.

The resolution in turbulence layer altitude and the maximum distance vary between the data sets depending on the asterism. In the phase B configuration, the minimum altitude resolution was varying between 1 and 1.5~km, while the maximum distance ranged from 18 to 28~km. This maximum distance is unaffected by the LGS asterism because of the cone effect and their relatively low altitude, but mostly determined by the off-axis distance of the NGS.

\begin{table}
	\begin{center}
		\begin{tabular}{|c|c|c|c|}
			\hline
			\# Asterism & A47 & A53 & A12 \\
			\hline
			Central $m_V$ & 11 & 10.9& 8.3\\
			\hline
			Sep~(") & 47.9 & 61.7 & 39.3 \\
			$m_V$ & 9.9 & 11.2 & 11.2\\
			\hline
			Sep~(") & 40.6 & 49.1 & 31.4 \\
			$m_V$ & 10.2 & 9.9 & 10.7 \\
			\hline
			Sep~(") & 53 & 56.8 & 51.5 \\
			$m_V$ & 8.7 & 9.8 & 10 \\
			\hline
		\end{tabular}
	\end{center}
	\caption{{Description of main asterism observed in \cana phase B. The columns indicate the CANARY asterism reference number, the separation (in arc-seconds) of each off-axis star to the central one and the V band magnitudes of each. See \cite{Vidal2014} for asterism illustration.}}
	\label{T:ast}
\end{table}

\begin{figure}
	\begin{center}
			\includegraphics[scale=.49]{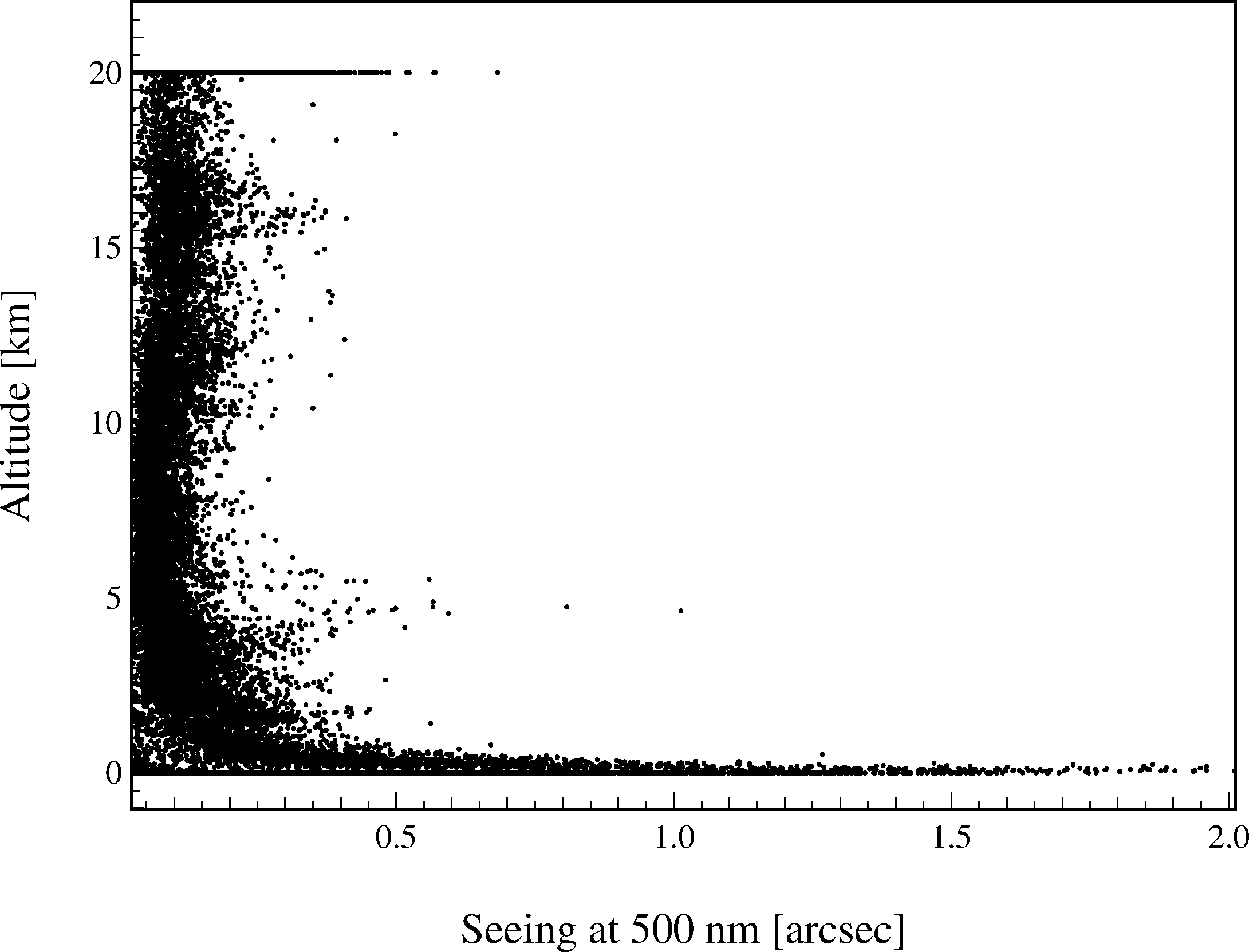}
			\caption{{Local seeing versus altitude. Each point corresponds to a result from the Learn\&Apply algorithm over about 4,500 processed data sets. Seeing and distance are unbiased from airmass. The plateau we observe at 20~km gathers all the altitude layers retrieved above this limit imposed by the tomographic geometry with \cana.}}
			\label{F:stat}
	\end{center}
\end{figure}

We present in Fig.\ref{F:stat} all the retrieved layers on-sky giving for each point their altitude layers versus their local seeing at 500~nm. Each sample is a result from the L\&A algorithm applied on a single data set, over 4,500 sets acquired by \cana in 2013. This figure shows the distribution of the turbulence along the altitude as a probability density. It makes a strong ground layer appear, which is mainly concentrated in the first kilometer. There is a significant strength of the turbulence between 1 and 5~km. We also find a group of high altitude layers, around 12~km spreading over 4~km. A strong layer is observed around 16~km, then the strength of the turbulence decreases at higher altitudes.

\begin{figure}
	\begin{center}
			\includegraphics[scale=.49]{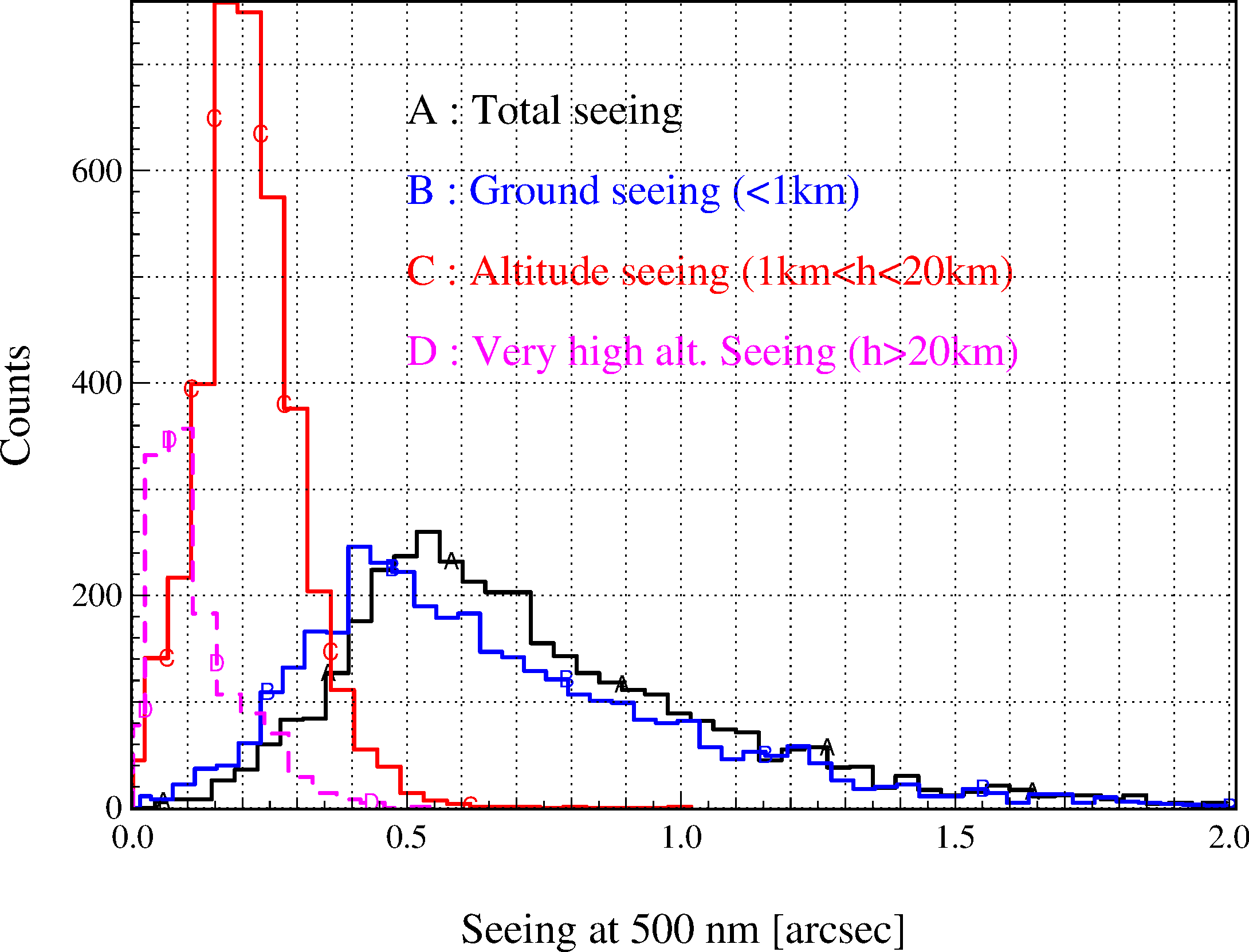}
			\caption{{Seeing histograms at 500~nm. Altitude layers are considered up to 1~km, which corresponds to the minimal tomographic resolution accessible with asterism observed using \cana \citep{Vidal2014}. Seeing is unbiased from airmass.}}
			\label{F:histo}
	\end{center}
\end{figure}

We present in Fig.~\ref{F:histo} histogram of the seeing for the ground layer~(altitude below 1~km), for the altitude layer and for the total atmosphere. The ground is defined by the sum of the contributions of the layers between 0 and 1~km. We get an altitude seeing that is relatively stable, around 0.21" median with a standard deviation of 0.09", while at the ground, the median seeing reaches 0.59" with a much larger standard deviation of 0.34". The median total seeing is 0.66" with a standard deviation of 0.33".

We conclude that the variation of seeing during our nights was mostly dominated by the variation of the ground layer strength. The seeing in altitude remains quite constant during the night while the ground evolves quickly and strongly. It changes the weights between altitude and ground layers in terms of strength. This introduces an additional error propagating through the MOAO reconstructor that can be mitigated only by updating this as often as possible. We note the GLAO reconstructor, that averages directly off-axis WFS measurements~(see Eqs.~\ref{E:Rglao} to~\ref{E:RglaoTT}), is not required to be updated.

Above 20~km, we are not able to identify the altitude layers properly. In Fig.~\ref{F:stat}, we have gathered all layers retrieved above 20~km at this maximal altitude. We can determine that we have around 0.089" seeing in median above 20~km, that corresponds to 240~nm on a 4.2~m telescope. However, we observe very high altitude layers only 34\% of the time in all the phase B observations. It means such processes, quite strong but rare, exist above the WHT and directly impact the tomographic performance of \cana.

\subsection{\cana on-sky results}
\label{SS42}

\begin{figure}
	\begin{center}
		\includegraphics[scale=0.49]{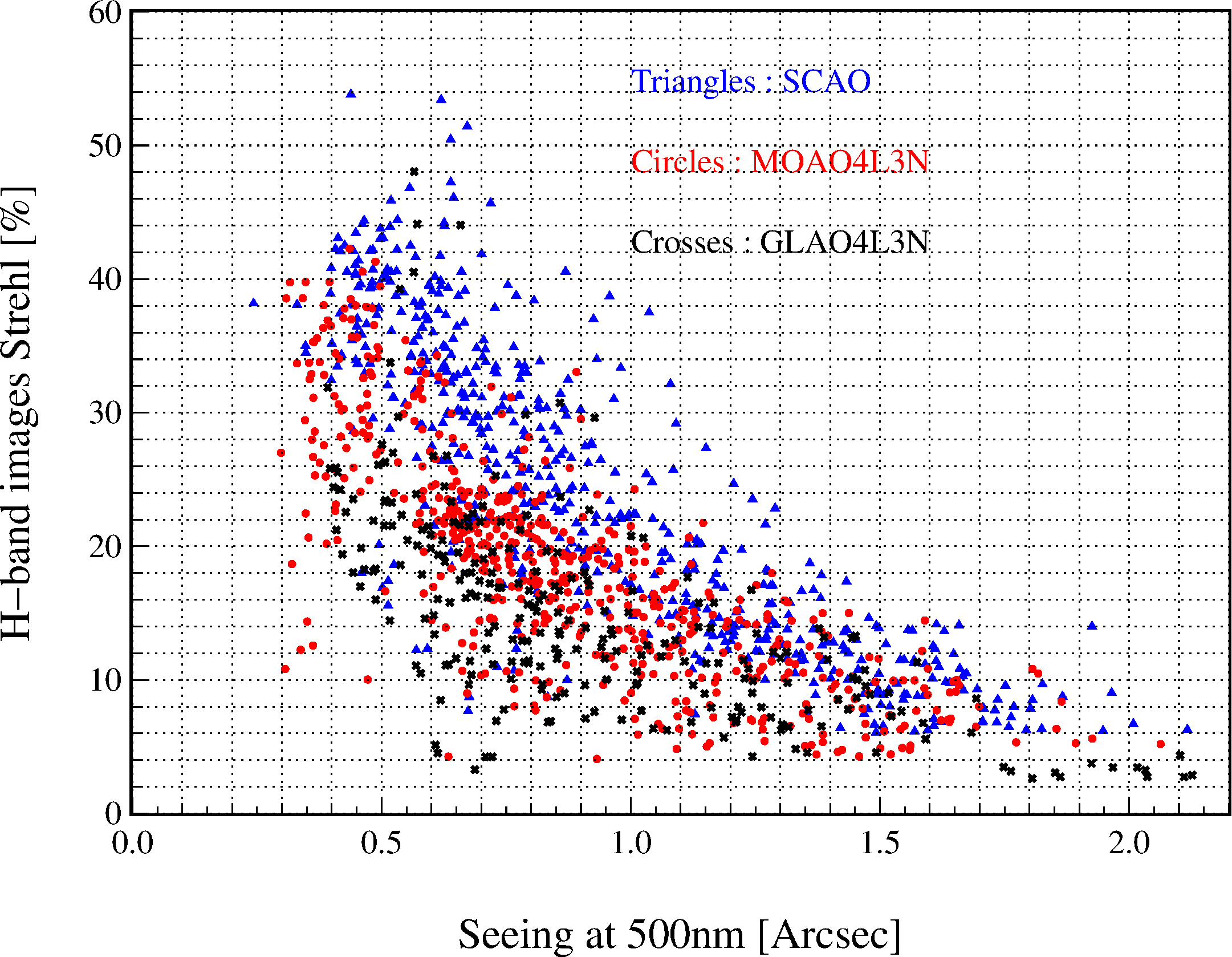}
		\caption{ {On-sky H-band image SR measured with \cana and \textsc{Camicaz} during phase B~(2013). We have concentrated results on SCAO, MOAO and GLAO modes with a configuration based on 3 NGS and 4 LGS.}}
		\label{F:sr}
	\end{center}
\end{figure}

We report in Fig.~\ref{F:sr} H-band image SR measured with \cana running in SCAO, MOAO and GLAO. We notice on SR three different performance regimes~: globally SCAO performs better than MOAO that performs better than GLAO.

Fig.~\ref{F:sr} makes a large scattering of results in any modes appear. In SCAO, this scattering involves mainly the wind speed variability and stars magnitude difference between asterisms.

In MOAO and GLAO, the scattering is mainly due to the turbulence profile variability with time. For a given seeing, the turbulence could be characterized by very different layer relative strengths.  Since GLAO is compensating the ground layer only, for a given seeing, the SR will drop down if the turbulence in altitude become stronger.Conversely, for the same seeing, the SR will increase if the ground layer dominates the turbulence.

In MOAO, this scattering is less expanded than in GLAO, as it is confirmed by Table~\ref{T:perf}, and this is what we should expect from a MOAO system~: if the tomographic reconstruction is properly achieved, the MOAO system must be robust to turbulent profile variability and provides the best reachable correction for a given $\cnh$ profile.

In Table~\ref{T:perf}, we give median values of SR for various ranges of seeing to illustrate what we have previously said. For each range, we get the best performance in SCAO and the worst in GLAO. In addition, results obtained with GLAO are more variable than MOAO and SCAO.

Below 0.6" of seeing, we were working with good to exceptionally good conditions. The turbulence was not dominated by the ground layer and the tomographic reconstruction has operated well in comparing MOAO and GLAO. The difference between SCAO and MOAO comes essentially from the tomographic resolution we have. At this level of seeing, it becomes very important to identify accurately the $\cnh$ profile to take into account weak layers, that are no more dominated by the most probable layers identified in Fig.~\ref{F:stat}.

Between 0.6 and 0.8", we were observing in nominal observation conditions. Table~\ref{T:perf} gives nominal performance of \cana on-sky. For seeing larger than 0.8 arcsec, the SR is dropping down drastically and for very bad conditions, seeing larger than 1.2 arcsec, SCAO, MOAO and GLAO perform at similar levels. It directly comes from our previous discussion on the seeing distribution~: when the seeing is so bad, it means the turbulence is dominated by the ground layer. The tomographic reconstruction can not provide a real improvement on performance as it does for better conditions.
\begin{table}
	\begin{center}
		\begin{tabular}{|P{1.5cm}|c|c|c|}
			\hline
			Seeing ["] & SR SCAO & SR MOAO & SR GLAO \rule[-2pt]{0pt}{12pt}\\
			\hline
			< 0.6     & 37.3\% $\pm$ 0.7 & 30.5\% $\pm$ 1.2& 22.1\% $\pm$ 1.8  \rule[-2pt]{0pt}{12pt}\\
			\hline
			0.6 - 0.8 & 30.1\% $\pm$ 0.7 & 21.4\% $\pm$ 1.0 & 17.1\% $\pm$ 1.5 \rule[-2pt]{0pt}{12pt}\\
			\hline
			0.8 - 1.2 & 20.6\% $\pm$ 0.5 & 15.6\% $\pm$ 0.9 & 11.0\% $\pm$ 1.5 \rule[-2pt]{0pt}{12pt}\\
			\hline
			> 1.2     & 11.5\% $\pm$ 0.3 & 10.2\% $\pm$ 1.3 & 9.0\% $\pm$ 1.9  \rule[-2pt]{0pt}{12pt}\\
			\hline
		\end{tabular}
	\end{center}
	\caption{{Median values of H-band image SR in SCAO, MOAO and GLAO, with a 3~NGS/4~LGS configuration, for different ranges of seeing. Error bars are given at one sigma and the seeing is given at 500 nm and unbiased from airmass.}}
	\label{T:perf}
\end{table}

\subsection{Statistical comparison with the IT and TS methods}
\label{SS43}

\begin{figure}
	\begin{center}
		\includegraphics[scale=0.49]{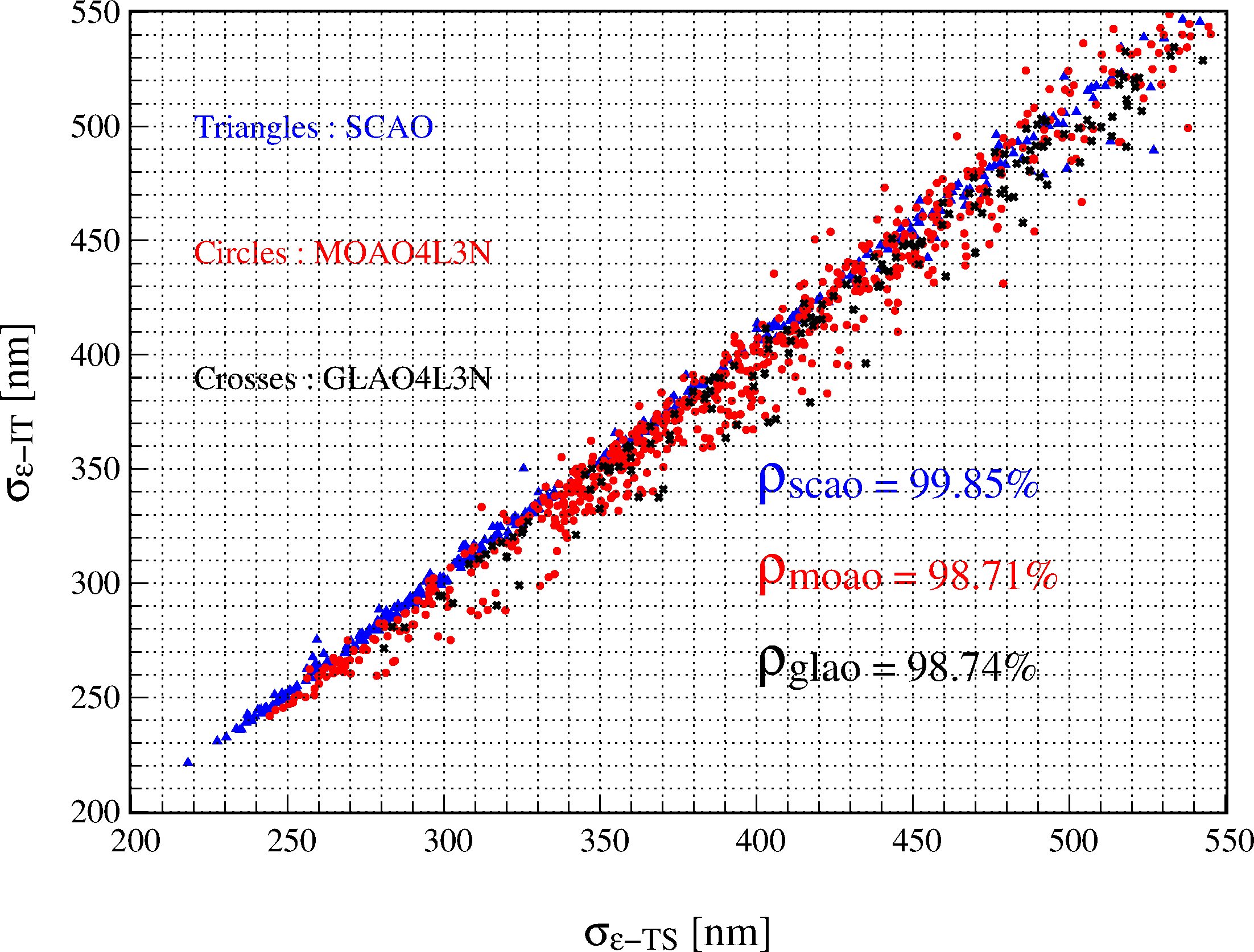}
		\caption{ { Rms value of the residual phase computed analytically from the IT method versus the empirical one got from the TS method. Less than 1\% of the points are off ($> 1\sigma$), most of them in MOAO. Coefficients given at the right-down corner are the estimation of the correlation factor~(Pearson coefficient) between analytic SR and image SR. SCAO={$\bigtriangleup$}, MOAO=$\bigcirc$, GLAO=$\times$. }}
		\label{F:comparVar}
	\end{center}
\end{figure}

In Fig.~\ref{F:comparVar}, we present the residual WF error $\sig{$\varepsilon_\text{IT}$}$~(see Eq.~\ref{E:DTI}) versus $\sig{$\varepsilon_\text{TS}$}$~(see Eq.~\ref{E:DMTS}). By comparing these two quantities, we quantify the accuracy of our error breakdown proposed in Sect.~\ref{S3}. We get about 99~\% of correlation in all modes and less than 1~\% of the total points are further than one sigma from the y=x line. We get a larger scattering in MOAO and GLAO since the error breakdown evaluation depends directly on the $\cnh$ profile estimation. We accumulate more potential sources of error.

{A posteriori}, the very high correlation between analytic calculations and measurements shows all the assumptions we have made in Sect.~\ref{S3} are not limiting results at this scale of atmosphere compensation. 

On top of that, Fig.~\ref{F:comparVar} illustrates the identification of the turbulence has been done properly, with enough accuracy and precision to get those fitting between calculations and observations. We do not observe any bias whatever the range of residual WF error as well. We are thus able to determine the turbulence characteristics, at a same level of accuracy, whatever the seeing conditions.

These results make us confident in future developments for preparing the future MOS design simulations, and especially for \textsc{MOSAIC} the next multi-object instrument proposed for the E-ELT.\\

\begin{figure}
	\begin{center}
		\includegraphics[scale=0.49]{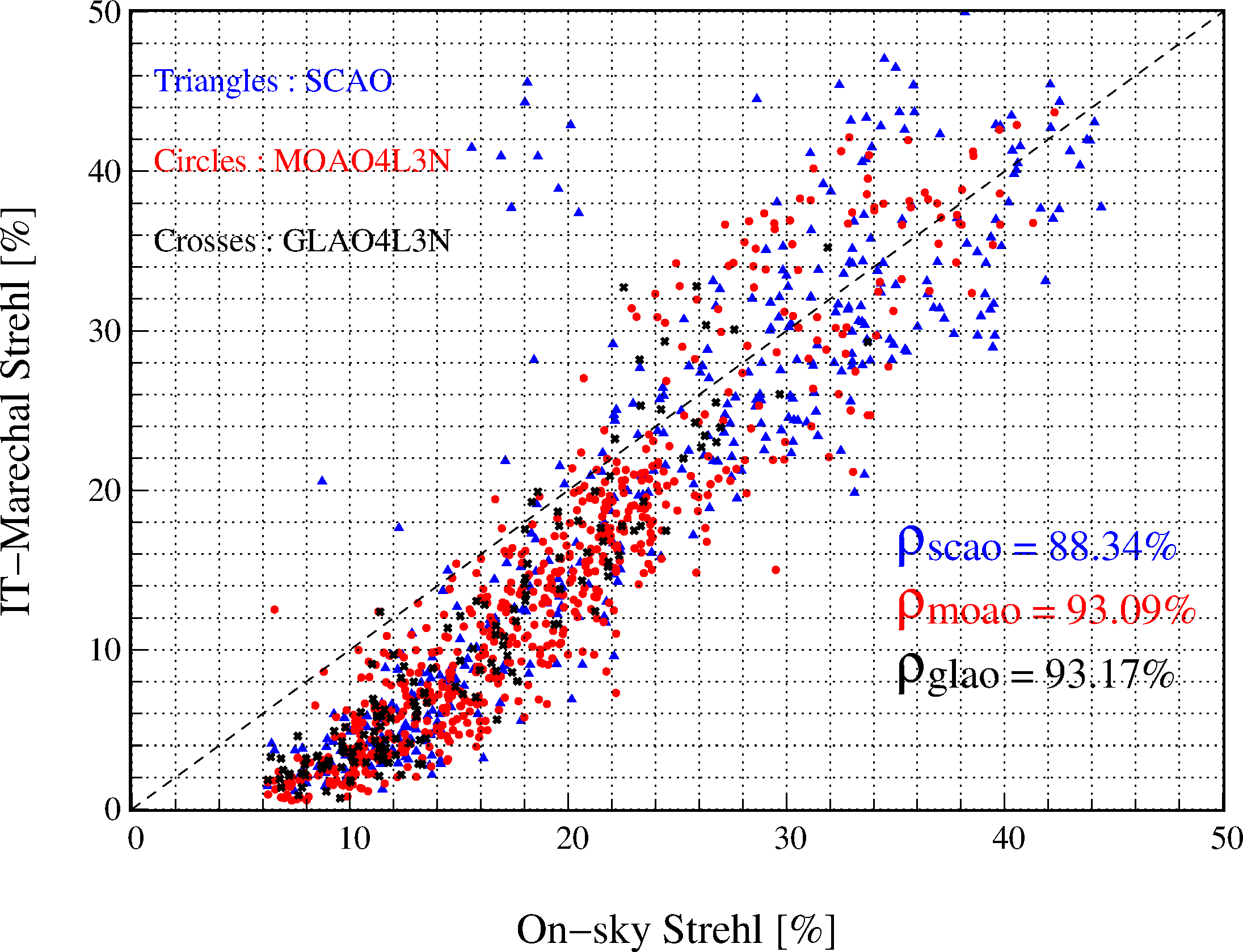}
		\caption{ { Analytic SR estimated using Mar\'echal approximation from $\sigdeux{$\varepsilon_\text{IT}$}$ versus H-band image SR. Coefficients given at the right-down corner are the estimation of the correlation factor~(Pearson coefficient) between analytic SR and image SR. SCAO={$\bigtriangleup$}, MOAO=$\bigcirc$, GLAO=$\times$.}}
		\label{F:SRmar}
	\end{center}
\end{figure}

\begin{figure}
	\begin{center}
		\includegraphics[scale=0.49]{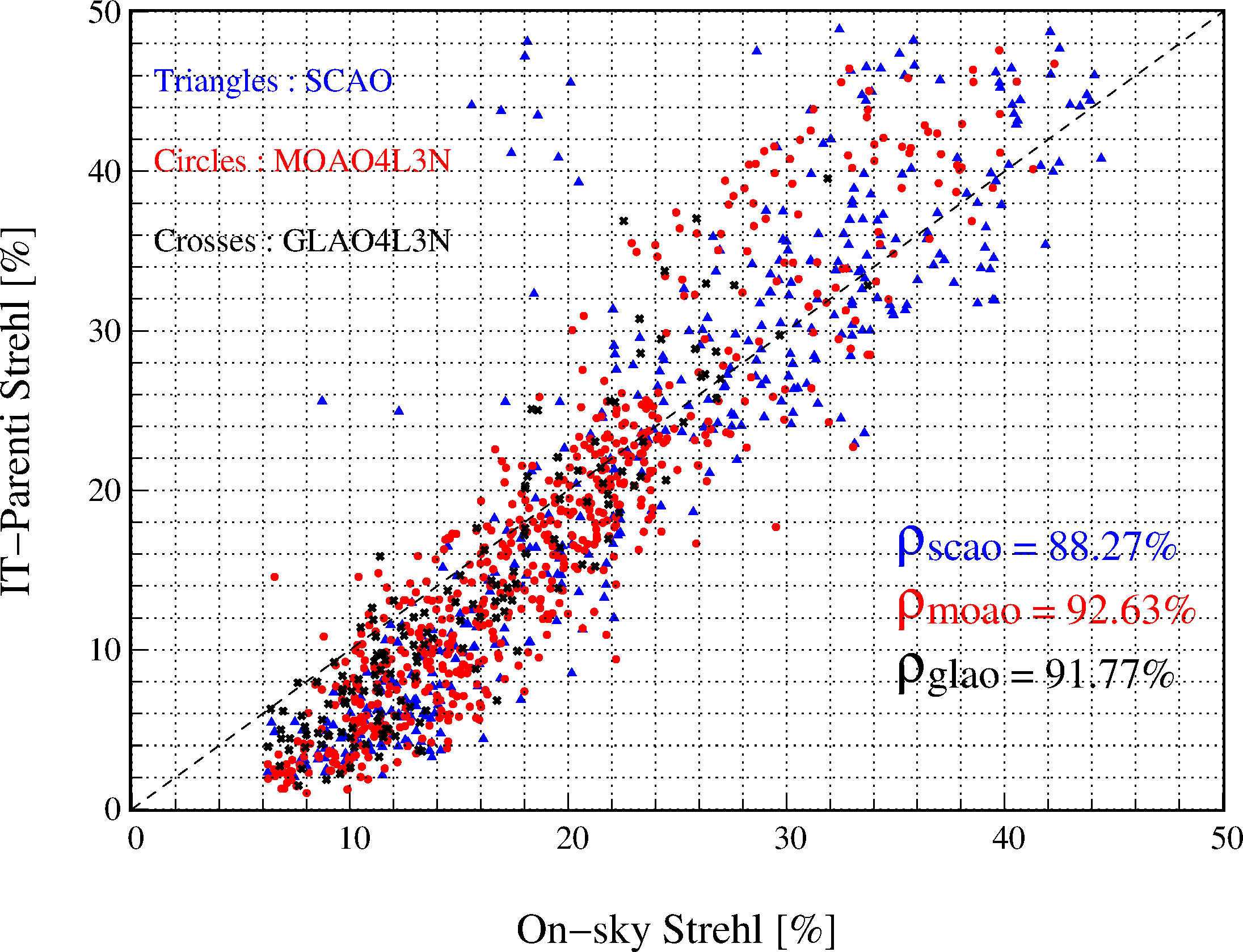}
		\caption{ {Estimate of SR using Eq.~\ref{E:SR} from  $\sigdeux{$\varepsilon_\text{IT}$}$ versus image SR. Coefficients given at the right-down corner are the results the estimation of the correlation factor~(Pearson coefficient) between analytic SR and image SR. SCAO={$\bigtriangleup$}, MOAO=$\bigcirc$, GLAO=$\times$.}}
		\label{F:SRPar}
	\end{center}
\end{figure}

Figs.~\ref{F:SRmar} and~\ref{F:SRPar} show the H-band SR estimated using respectively Eqs.~\ref{E:SRmar} and~\ref{E:SR}, versus the sky image SR. Both of Mar\'echal and Parenti approximations are using $\sigdeux{$\varepsilon_\text{IT}$}$ in Eq.~\ref{E:DTI} to estimate the SR.
The relation between the SR and the residual WF variance is not bijective~: for a given variance $\sigdeux{$\varepsilon$}$, several values of SR are possible. For instance, the tip-tilt modes make the PSF core spread. For a given residual variance, if there is more residual tip-tilt in a situation than in another, the first one will comes with a worse SR. The Parenti formulation proposed in Eq.~\ref{F:SRPar} tries to take into account this modal weighting, but the real relation between SR and variance cannot be given by a simple formula.

In Fig.~\ref{F:SRmar}, we observe a correlation between analytic and image SR for  SR larger than 20\%, but find a  bias for lower values of image SR. For both methods, we get very similar Pearson coefficient, but the Parenti approximation makes the under-estimation of low SR disappear as shown in Fig.~\ref{F:SRPar}. The scattering we observe on Figs.~\ref{F:SRmar} and~\ref{F:SRPar} comes from the inaccuracy of Mar\'echal and Parenti approximations, since we do not observe such a scattering in Fig.~\ref{F:comparVar}.  



\section{Analysis of the joint NGS plus LGS tomography performed on-sky}
\label{S5}

We focus in this section on the \cana performance to target the improvement of the atmosphere compensation, achieved thanks to both tomographic reconstruction and LGS. We compare on-sky results acquired during what we call a {script}. A script is five data recordings of the same sequence made of three successive AO modes observation. A sequence lasts 45~s, 15~s per AO mode. This duration has been chosen to enable us to compare \cana performance, running in different AO modes, in similar atmospheric conditions.

We compare on-sky performance achieved by SCAO, GLAO and MOAO with and without LGS. We use the letters "N" and "L" following the AO modes to define respectively NGS and LGS sensors. For instance, "MOAO4L3N" means \cana was compensating the turbulence in MOAO mode using the four LGS and three NGS.

In the following, we will present three of those scripts, numbered 274, 292 and 275. In Table~\ref{T:scriptconditions}, we report observation conditions we have identified during those scripts. The seeing and outer scale values comes from the Learn\&Apply algorithm. 

The wind speed has been retrieved from the full width at half maximum of the slope measurement temporal auto-correlation function, after subtraction of the noise contribution. This method provides an integrated value of the wind speed along the turbulence profile.

These scripts have been selected because of their relatively close observing conditions. The observed asterism was A47 for scripts 274 and 275 and A53 for 292~\citep{Vidal2014}. 

\begin{table*}
	\centering
	\begin{tabular}{|c|c|c|c|c|}
		\hline
		& Script&  274 & 292 & 275 \rule[-2pt]{0pt}{12pt}\\
		\hline
		& mode 1& SCAO & SCAO & SCAO \rule[-2pt]{0pt}{12pt}\\
		AO modes & mode 2& MOAO 4L3N & MOAO 4L3N & MOAO 4L3N \rule[-2pt]{0pt}{12pt}\\
		& mode 3 & GLAO 4L3N& GLAO 4L3N & MOAO 3N \rule[-2pt]{0pt}{12pt}\\
		\hline
		           & $\R$ calibration & 21~h 23~m & 01~h 37~m & 21~h 23~m \rule[-2pt]{0pt}{12pt}\\
		Local time & Script start     & 21~h 55~m & 02~h 30~m & 22~h 03~m \rule[-2pt]{0pt}{12pt}\\
		           & Script end       & 22~h 01~m & 02~h 35~m & 22~h 08~m \rule[-2pt]{0pt}{12pt}\\	
        \hline
		Asterism &  & A47 & A53 & A47\rule[-2pt]{0pt}{12pt} \\
        \hline
		Airmass range &  & 1.086 - 1.079 & 1.035 - 1.038 & 1.076 - 1.069\rule[-2pt]{0pt}{12pt} \\           
		\hline
		            & Total    & 0.83 & 0.78 & 0.81\rule[-2pt]{0pt}{15pt}\\
		Seeing ["]  & Ground   & 0.70 & 0.56 & 0.68\rule[-2pt]{0pt}{12pt}\\
		            & Altitude & 0.35 & 0.46 & 0.36\rule[-2pt]{0pt}{12pt}\\
		\hline  
	                & Total    & 13.3 & 12.2 & 11.9\rule[-2pt]{0pt}{12pt} \\
	    $\lz$ [m]   & Ground   & 5.2 & 5.9  & 12.1  \rule[-2pt]{0pt}{12pt}\\
	                & Altitude & 19.3 & 17.9 & 11.7\rule[-2pt]{0pt}{12pt}\\
	    \hline
	    Wind speed [m/s] & Total & 3.0 & 2.9 & 3.1\rule[-2pt]{0pt}{12pt} \\
		\hline
		                & Mode 1   & 28.3 & 30.0 & 29.0\rule[-2pt]{0pt}{12pt} \\
		Image SR [\%] & Mode 2   & 21.0 & 16.0 & 18.1 \rule[-2pt]{0pt}{12pt}\\
		                & Mode 3   & 12.9 & 9.2  & 14.6\rule[-2pt]{0pt}{12pt}\\
		 \hline
	\end{tabular}    
	\caption{{Observing conditions including turbulence parameters and \cana H-band image SR during scripts 274, 292 and 275. Each parameter has been retrieved an averaged on 15 successive 15~s data sets. Wind speed retrieved from the FWHM of slopes temporal auto-correlation function.}}
	\label{T:scriptconditions}
\end{table*}

\subsection{Script 274~: SCAO/MOAO 4L3N/GLAO 4L3N}

\begin{figure*}
	\begin{center}
		\includegraphics[scale=.6]{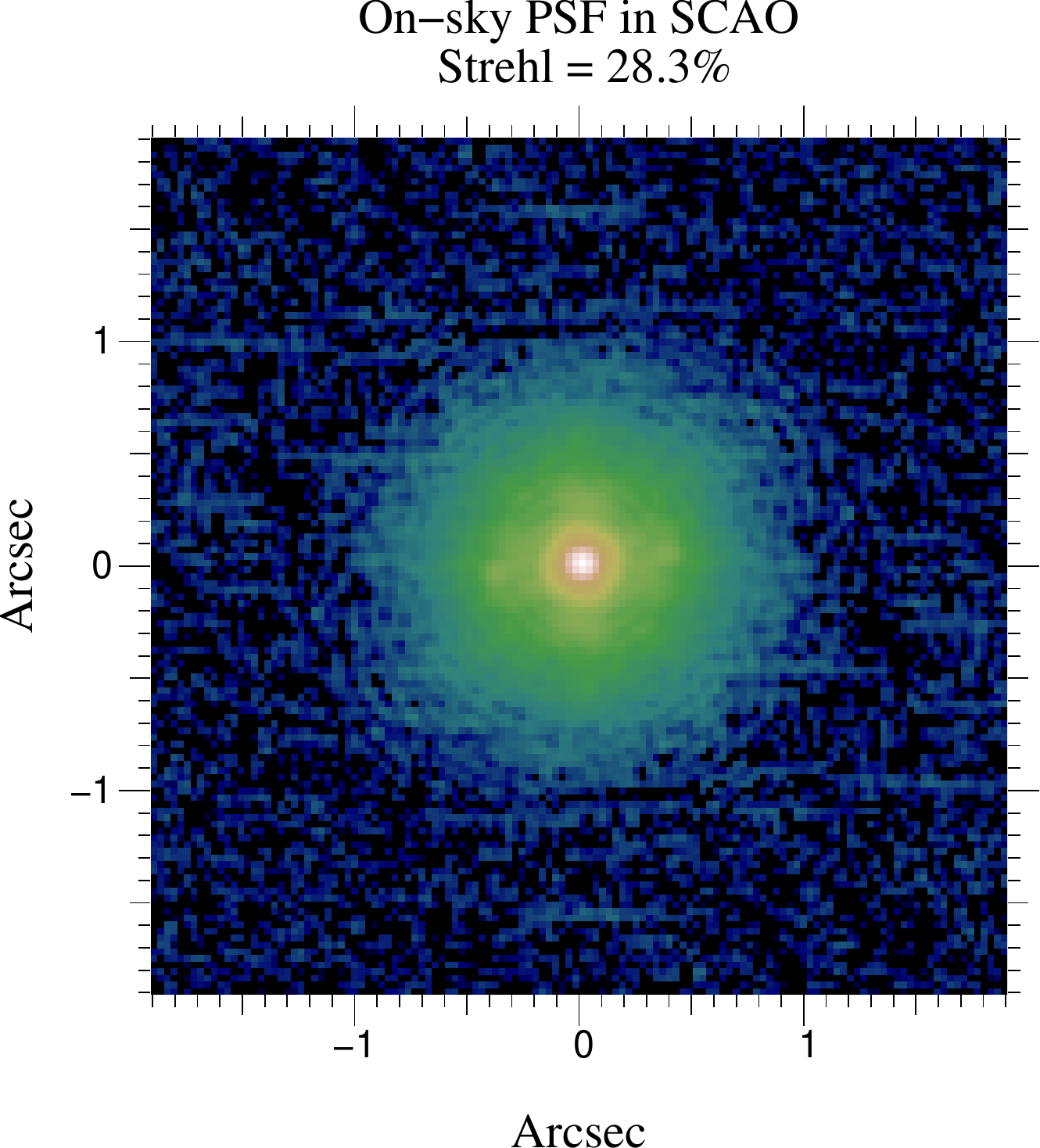}
		\includegraphics[scale=.6]{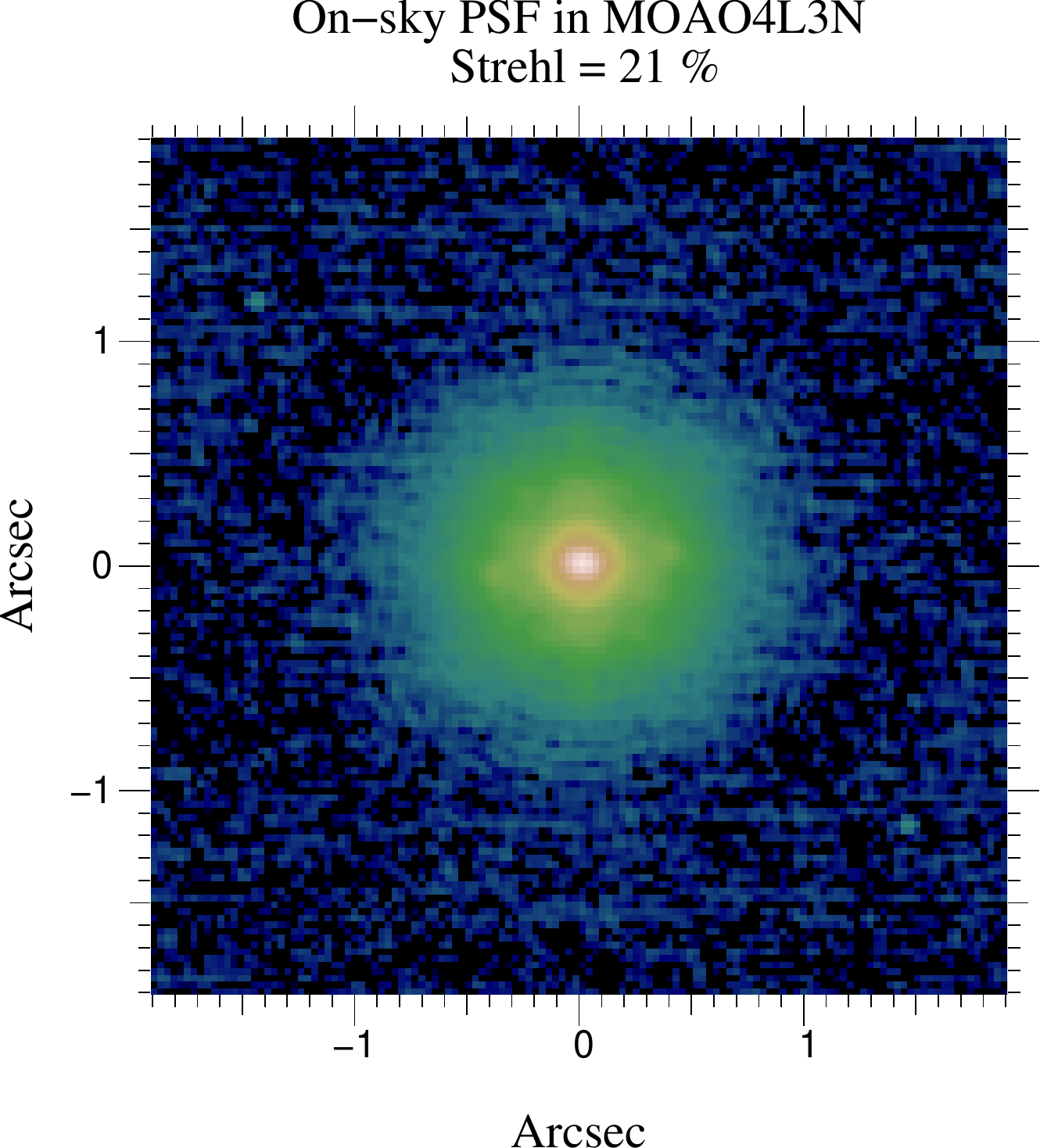}
		\includegraphics[scale=.6]{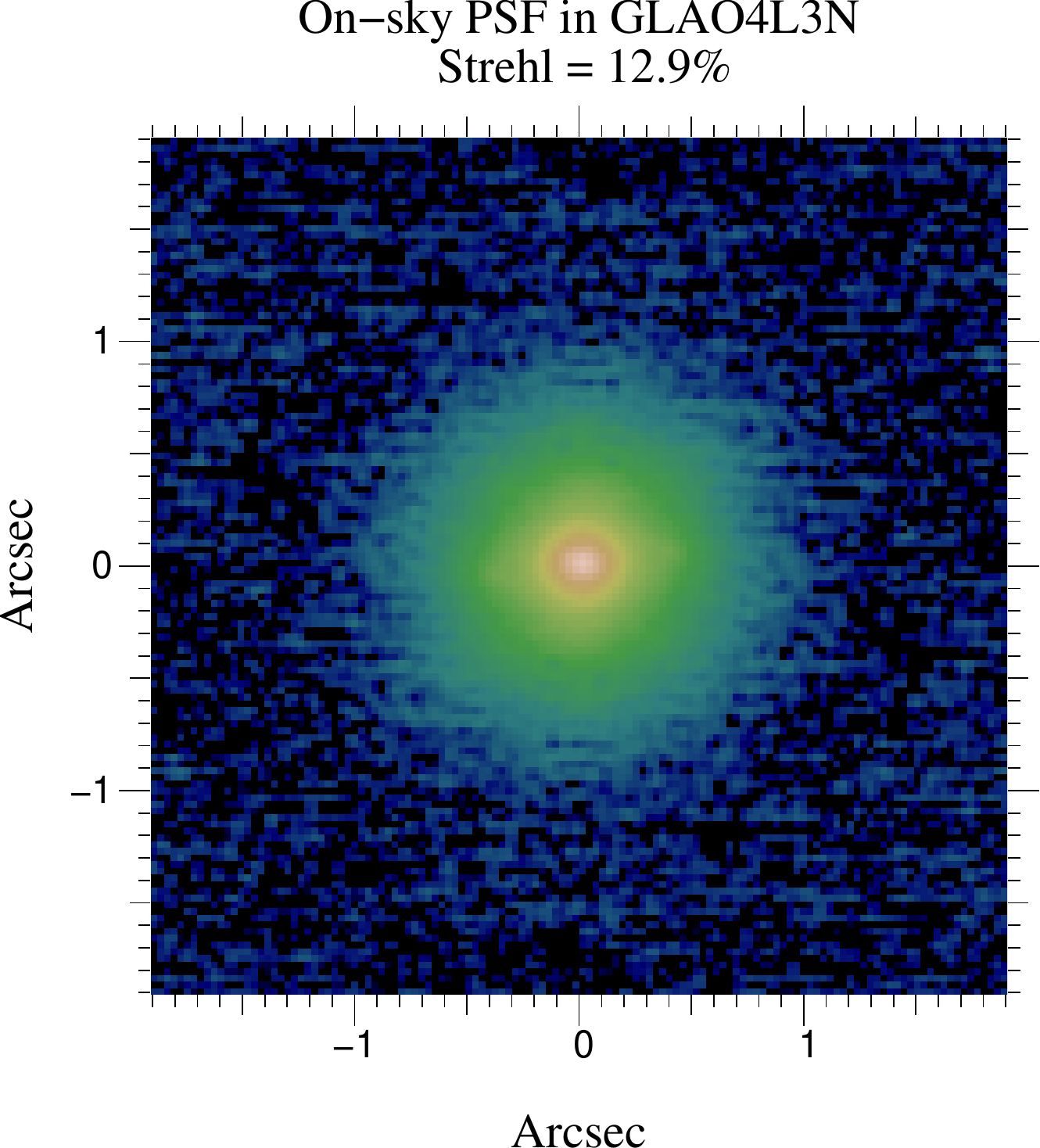}
		\caption{{Three AO compensated H-band PSF (log. scale) for SCAO, MOAO and GLAO. Each PSF is averaged on five IR images, not re-centred, acquired on the same AO mode during script 274, the September 13th 2013 night.}}
        \label{F:PSF}
	\end{center}
\end{figure*}

In Fig.~\ref{F:PSF}, we show three PSFs averaged over five H-band images acquired during script 274 in SCAO, MOAO 4L3N and GLAO 4L3N, the September 13th 2013 night. Thanks to the tomographic reconstruction, we make the SR improving from 12.9\% in GLAO to 21.0\% in MOAO, corresponding to a wave-front error reduction by an order of 190nm rms between the two modes, using Eq.~\ref{E:SRmar}. The best SR is obtained by SCAO with a value of 28.3\%.

\begin{figure}
	\begin{center}
		\includegraphics[scale=.49]{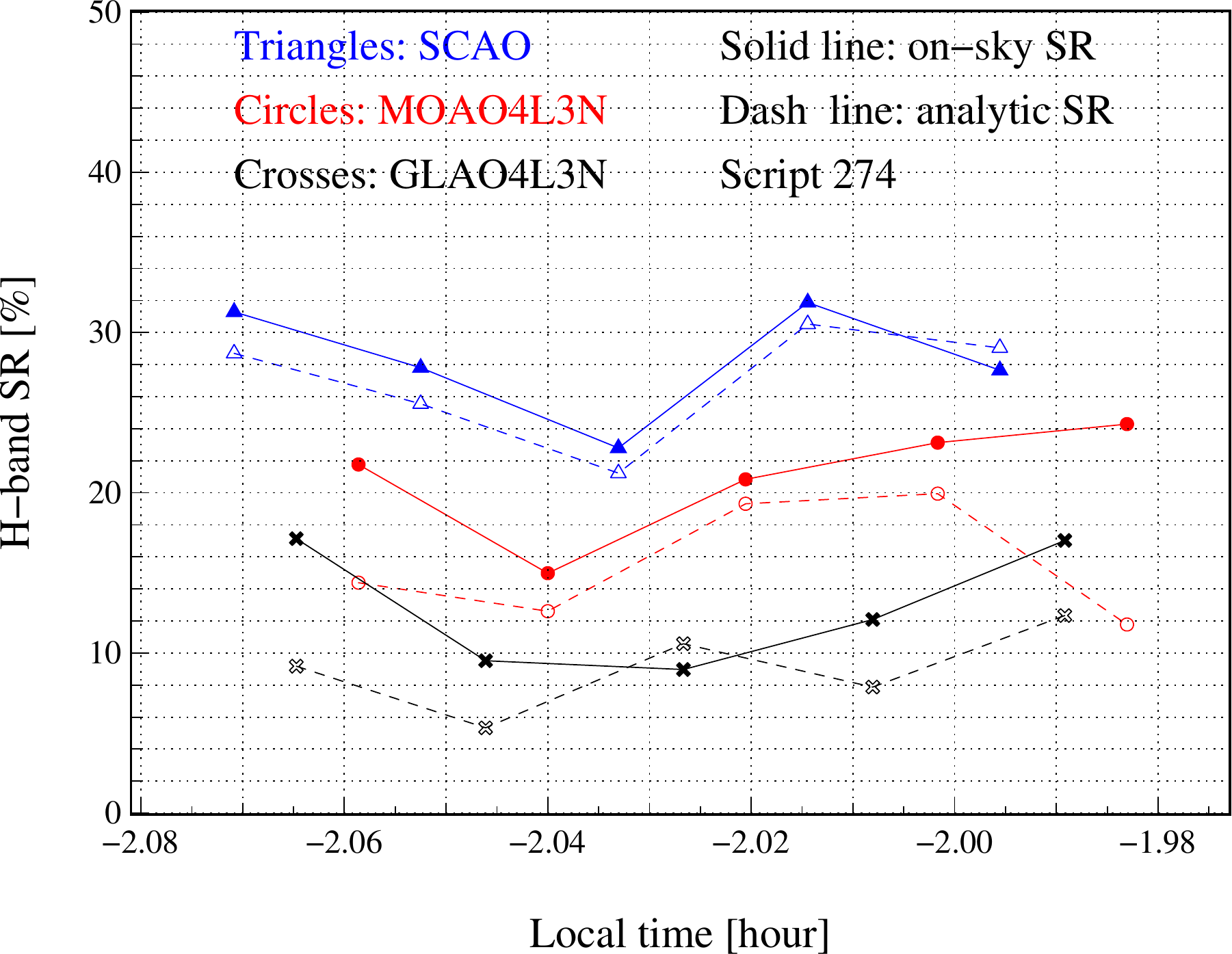}
		\caption{{Both IR image and analytic~(see Eq.~\ref{E:SR}) SR versus local time for script 274. SCAO={$\triangle$}, MOAO LGS/NGS=$\times$, MOAO 3 NGS=$\bigcirc$. In MOAO and GLAO, we have used the IT approach to get the residual phase variance, while in SCAO we have used the TS approach.}}
		\label{F:274_sr}
	\end{center}
\end{figure}

In Fig.~\ref{F:274_sr}, we highlight the evolution of both analytic and IR image SR with time. Analytic SR have been derived from Eq.~\ref{E:SR} using the residual phase variance developed in Eq.~\ref{E:DTI}~(IT method). The figure displays the five samples per AO mode, acquired successively and interlaced between each other.

By interlacing different AO modes, we minimize the impact of the possible turbulence non-stationarity during the script. If a particular event happens during the script, we should observe it with each mode if it lasts for at least 45~s. This ensures that we are able to compare AO modes with nearly similar observational conditions.

Figure~\ref{F:274_sr} shows that we are in the most able to reproduce analytically the SR in a satisfactory way. The evolution of the IR image SR with time is well reproduced, as the absolute value and the difference between modes. However, we have an under-estimation of the SR that becomes more severe when the SR becomes higher, especially for SCAO, thus implied by the Parenti approximation according to Fig.~\ref{F:SRPar}. Note from the same figure, the Mar\'echal approximation would give worse reproduction of low SR.

\begin{figure}
\begin{center}
\includegraphics[scale=.4]{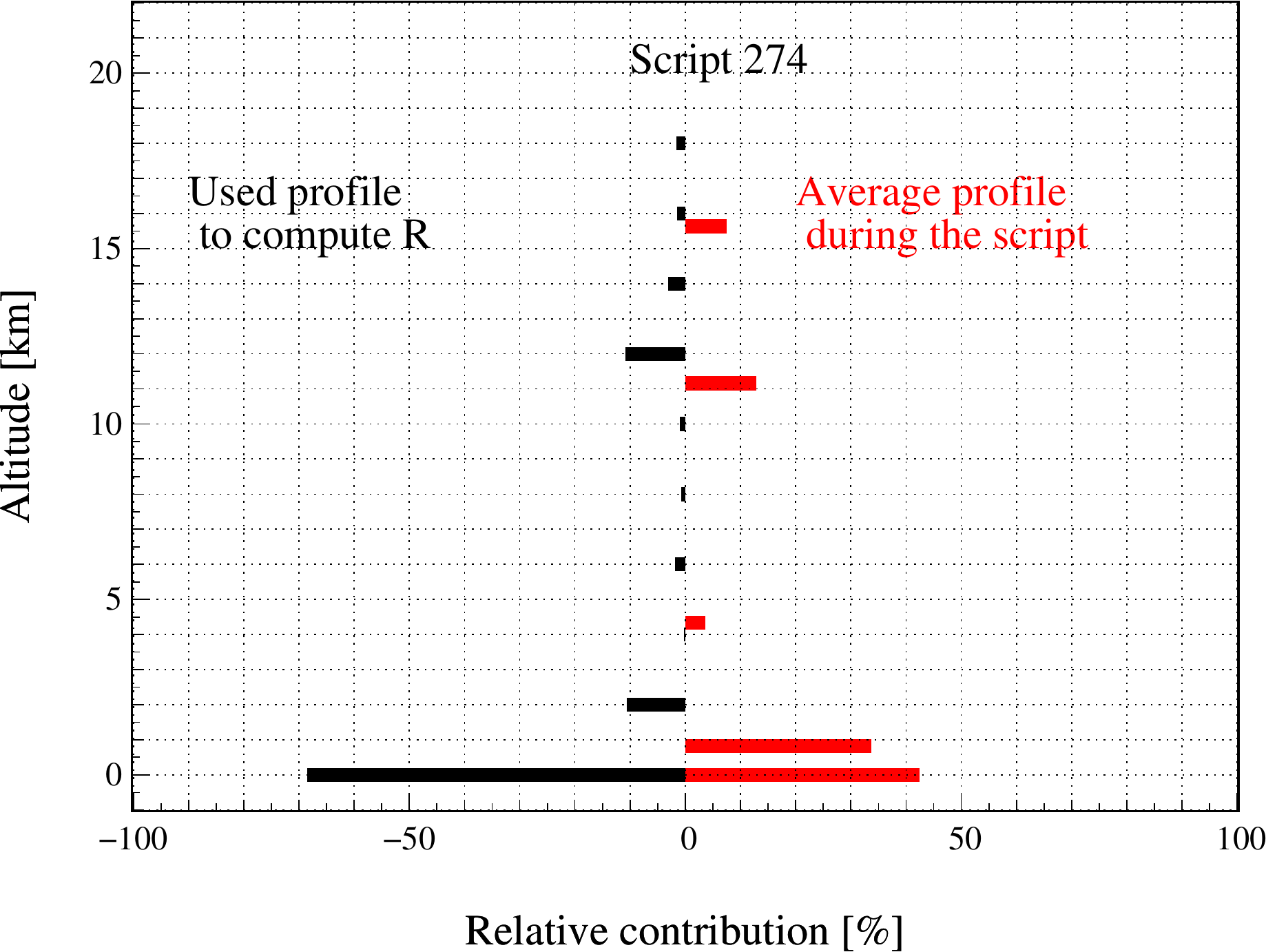}
\includegraphics[scale=.4]{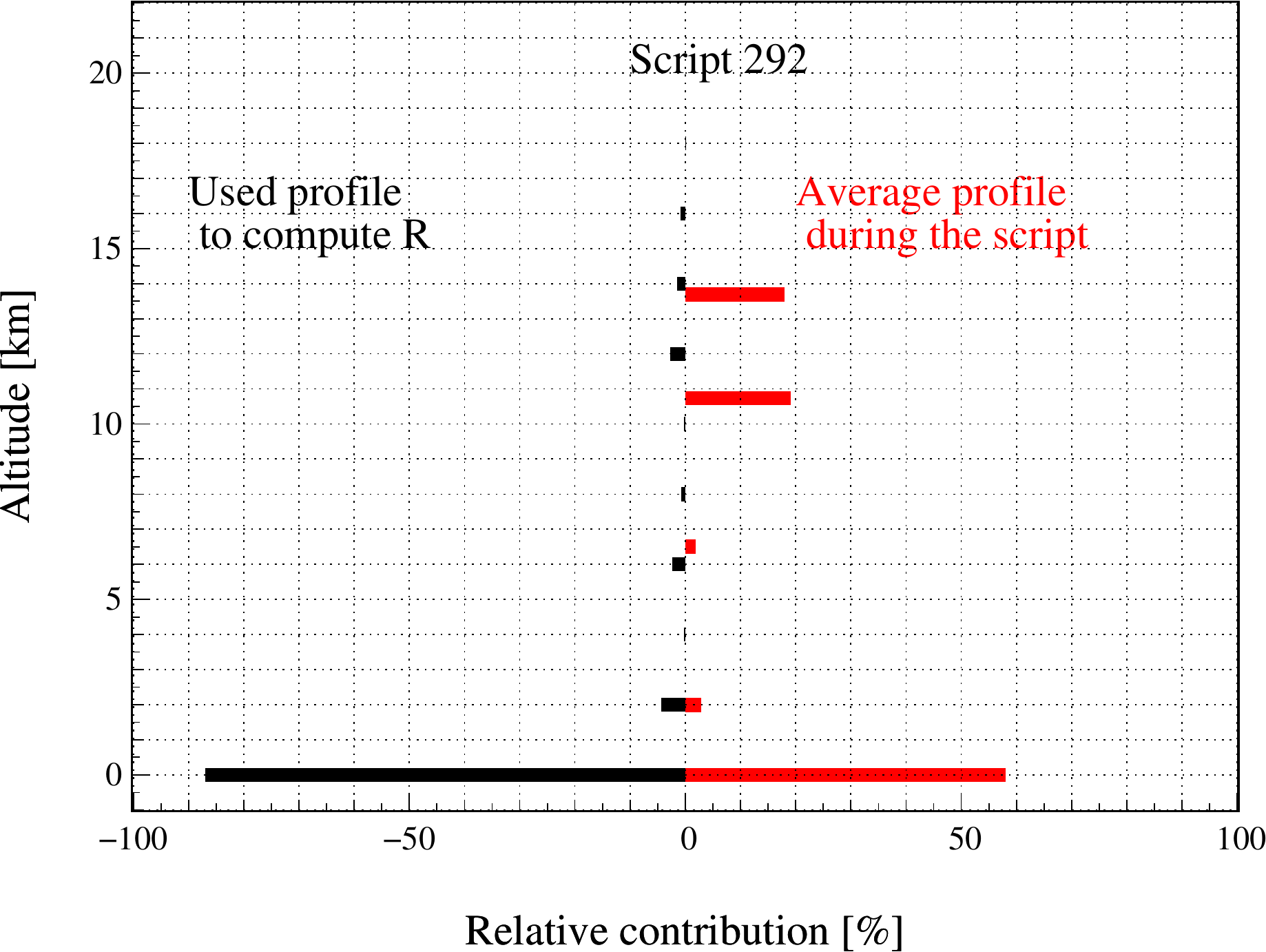}
\vspace{.25cm}

\includegraphics[scale=.4]{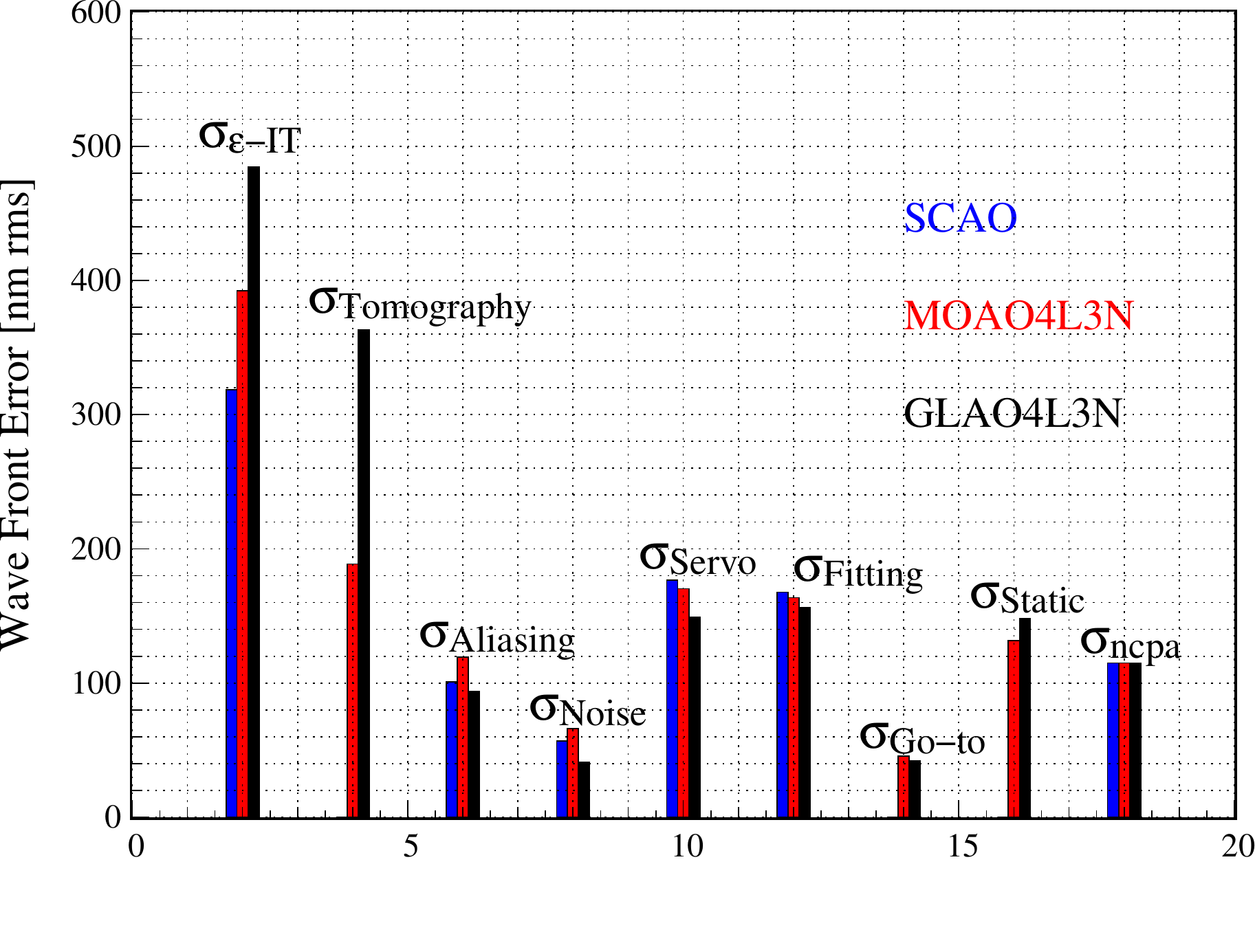}
\includegraphics[scale=.4]{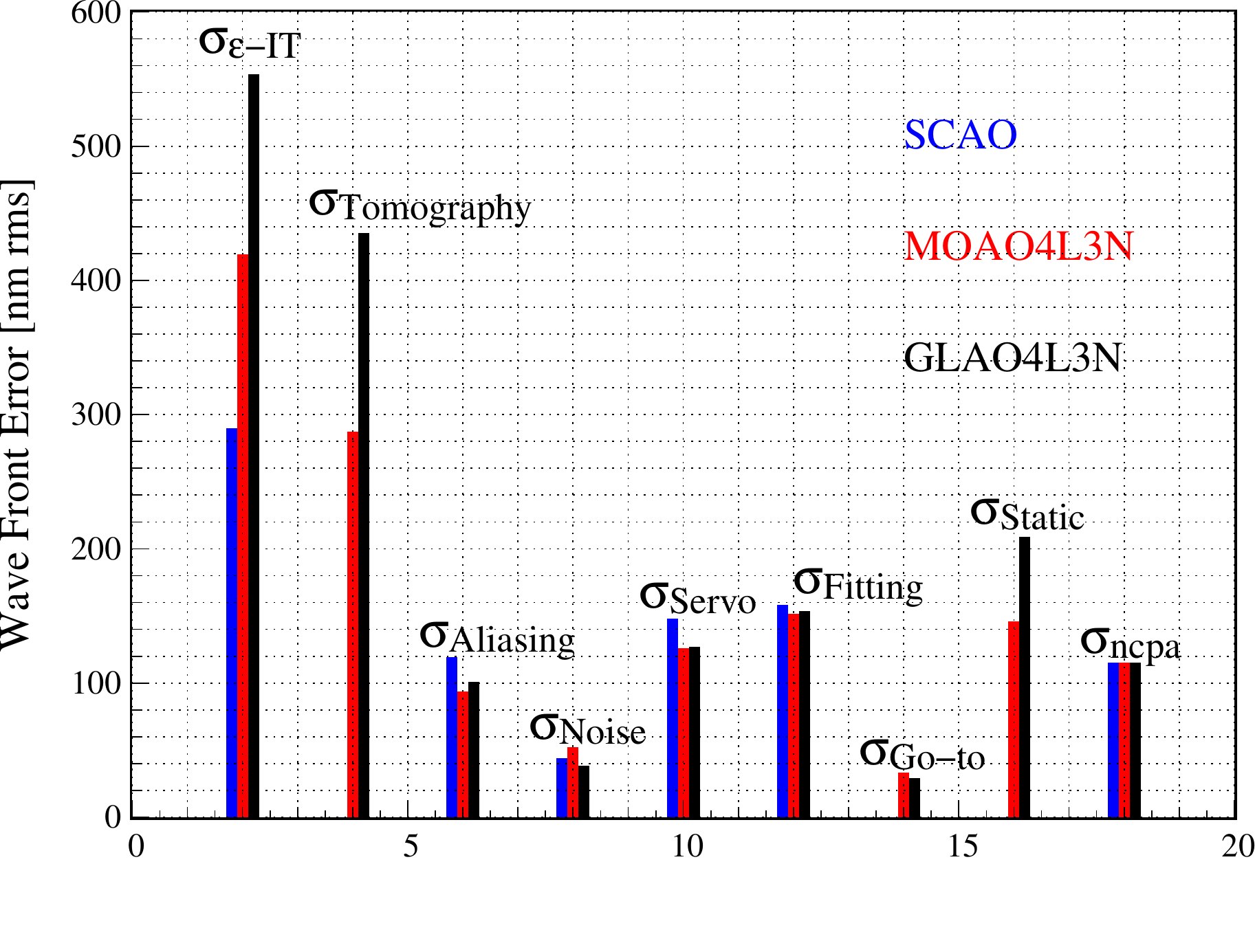}
\vspace{.25cm}

\includegraphics[scale=.4]{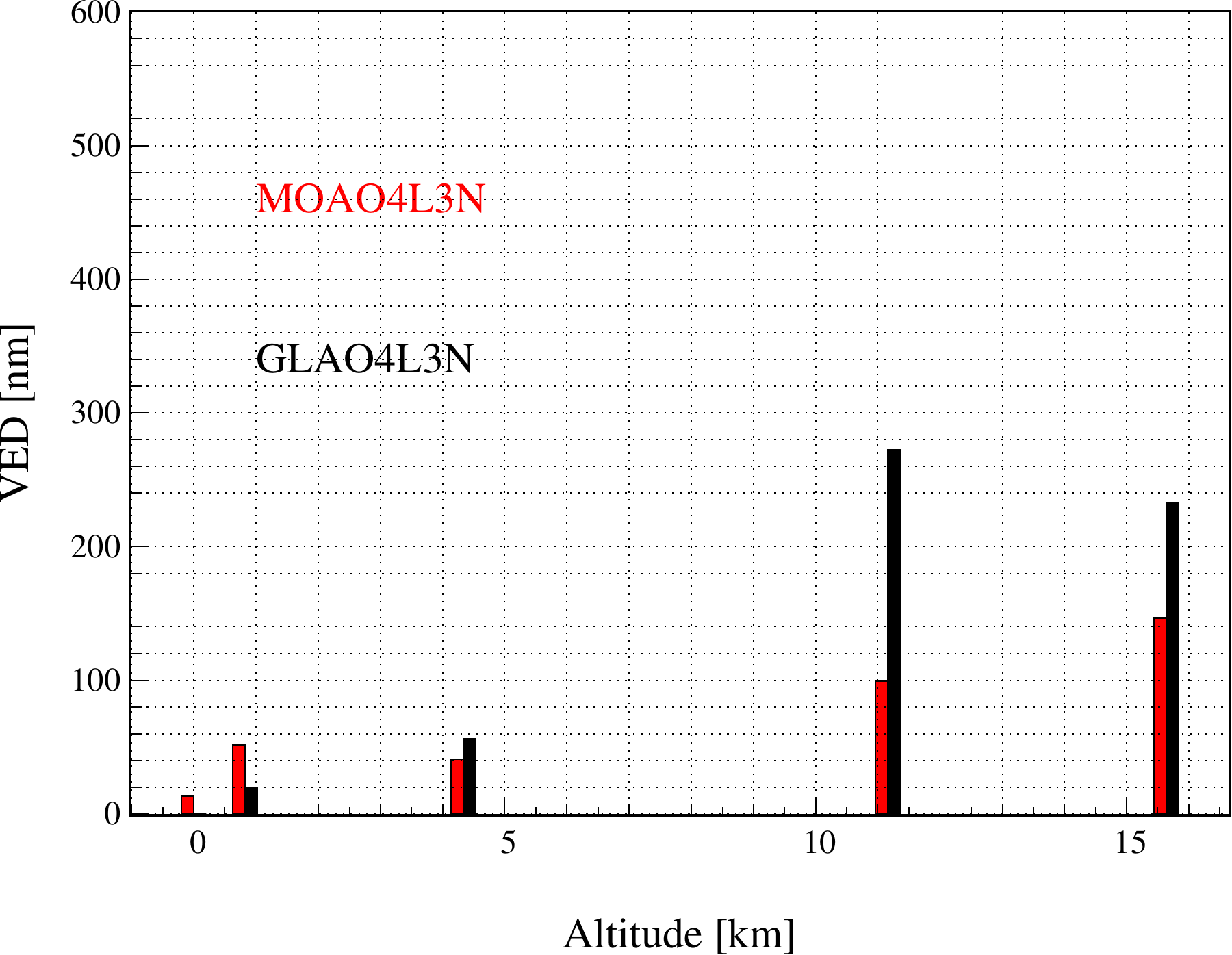}
\includegraphics[scale=.4]{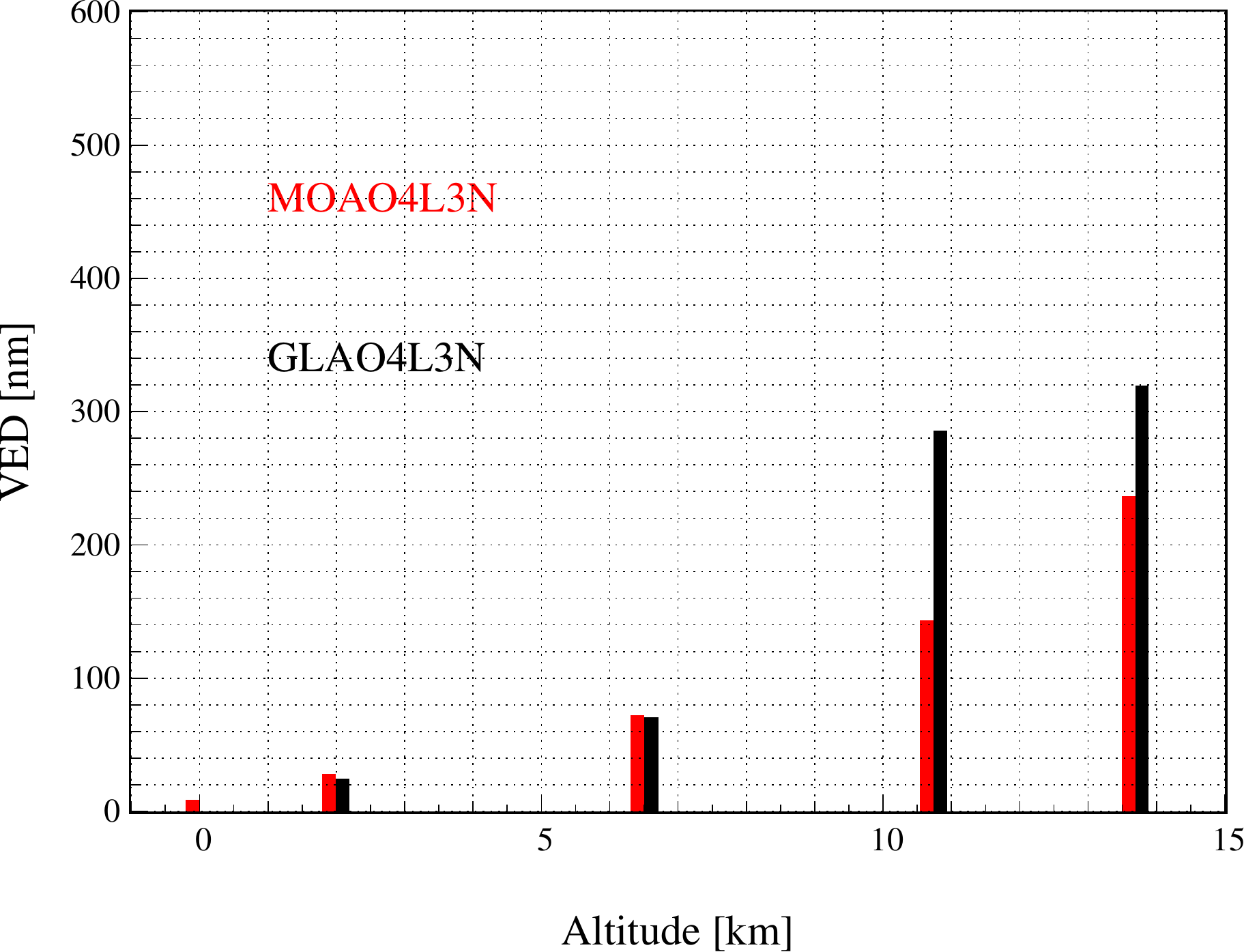}
\vspace{.25cm}

\caption{{\textbf{Left~:} Results for script 274. \textbf{Right~:} Results for script 292. \textbf{Up~:} Turbulent profiles: expected by the MOAO reconstructor (left), and averaged on the turbulent profiles identified on each data set acquired during the script (right). \textbf{Middle~:} Error breakdown decomposition for the three AO modes. Each variance has been averaged over the five realizations in the script. \textbf{Down~:} Tomographic VED for GLAO and MOAO. Each variance has been averaged over the five realizations. The quadratic sum of the VED values gives the tomographic error. }}
\label{F:274_prof}
\end{center}
\end{figure}

For further performance analysis, we present in Fig.~\ref{F:274_prof} top, in the left panel, a comparison between the turbulent profile on which the sky reconstructor has been calibrated and the post-retrieved one on the script 274 data sets. These results are from an average of fifteen profiles, each of which has been retrieved on five layers on each data set.
We obtain (by post processing) 75\% of turbulence at the ground~(below the first kilometer), against 70\% expected by the sky reconstructor (computed 20 min. before the script was run). In addition, the 11~km altitude layer observed during the script was quite well identified in the reconstructor. However, the observed 16~km altitude layer was partially identified thanks to the high altitude layers, between 14 and 18~km, as expected by the reconstructor. Finally, it seems that our MMSE reconstructor was still relevant for the turbulence profile encountered during the script.

In the middle, left panel of Fig.~\ref{F:274_prof}, we present the full error breakdown for each AO mode. It clearly highlights the improvement coming from the tomographic reconstruction~: the term $\sigdeux{Tomography}$ fell from 365~nm (GLAO) to 185~nm (MOAO), while the other terms were very similar. The difference between SCAO and MOAO is also explained by the tomographic error and the static terms as well. 

The fluctuations of the fitting, aliasing, servo-lag and noise errors we observe are relatively small between AO modes and mainly due to the variance on the seeing estimation. They also depend on the system temporal transfer function including the loop gain and the reconstructor used. 

In both MOAO and GLAO, we get 140~nm of static errors, despite the calibrations that have been done before the script~(see Sect.~\ref{SS:static}). It shows these calibrations are mandatory to operate the turbulence tomographic compensation properly.

In a further analysis of the tomographic error, we show the tomographic VED in Fig.~\ref{F:274_prof} bottom, left panel. It becomes  clear that the MOAO reconstructor allows better compensation of the altitude layers than the GLAO reconstructor. Note we reduce the WF error by 170~nm on the 11~km layer by using MOAO. However, the GLAO reconstructor is better than the MOAO one at compensating the very low altitude layers, below 2~km, near to be the tomographic resolution with this asterism. This is due to the robustness of MMSE reconstructor~: the more layers it predicts, the less efficiently it is able to compensate individual layers~\citep{Gendron2014}. This is for the same reason that the GLAO reconstructor is the only one to get 0~nm rms at 0~km, predicting this unique layer.

\subsection{Script 292~: SCAO/MOAO 4L3N/GLAO 4L3N}

We discuss here the results from script 292. In the same way as script 274, we compared SCAO, MOAO and GLAO, but with  a mis-calibration of the MMSE reconstructor. We illustrate in Fig.~\ref{F:274_prof} the turbulence profile used to compute the MOAO reconstructor, compared to the one retrieved during the script 292. According to Table~\ref{T:scriptconditions}, there was  one hour between the calibration and the script. During this time, strong altitude layers were appeared at 10.5 and 13.5~km, accounting together for 35~\% of the total turbulence and these are not expected by the MOAO reconstructor.

The global observation conditions, between script 292 and 274, was quite similar as given in Table~\ref{T:scriptconditions}. In addition, Fig.~\ref{F:274_prof} allows us to compare error breakdown between these two scripts, to highlight that we obtained the same level on each error term except for tomography. We get 285~nm rms during script 292 against 185 during script 274 for MOAO. These values are comparable since the turbulence profiles during observations, as shown in Fig.~\ref{F:274_prof}, were largely similar, with 60-70\% of ground turbulence and the rest of the energy distributed into two altitude layers.
The 215 nm rms difference in tomographic error is thus due to the mis-calibration of the tomographic reconstructor. It has led to a decrease of the IR image SR according to Table~\ref{T:scriptconditions}. 

Vertical error distribution plots given in Fig.~\ref{F:274_prof} also illustrate that the MOAO reconstructor has been mis-calibrated. However it still performs much better than the GLAO one, especially at high altitudes. For instance at the 11~km layer, we get similar performance in GLAO during the two scripts 274 and 292, with 275~nm and 285~nm of error respectively. We underline the tomographic errors of MOAO in script 292 are very significant for the two high altitude layers at 11 and 13.5 km with 140nm and 235 nm respectively. Much poorer performance is shown when compared to the high altitude layers of script 274. 

The MOAO reconstructor has to be calibrated several times each night. It is thus mandatory to propose innovative techniques to compute and update a MMSE reconstructor for large number of degrees of freedom in order to prepare the E-ELT. Fast computation and data transfer are required. Several works are on-going on that, in particular using GPUs~\citep{Gratadour2013c,Gratadour2012}.

\subsection{Script 275~: SCAO/MOAO 4L3N/MOAO 3N}

\begin{figure}
	\begin{center}
		\includegraphics[scale=.49]{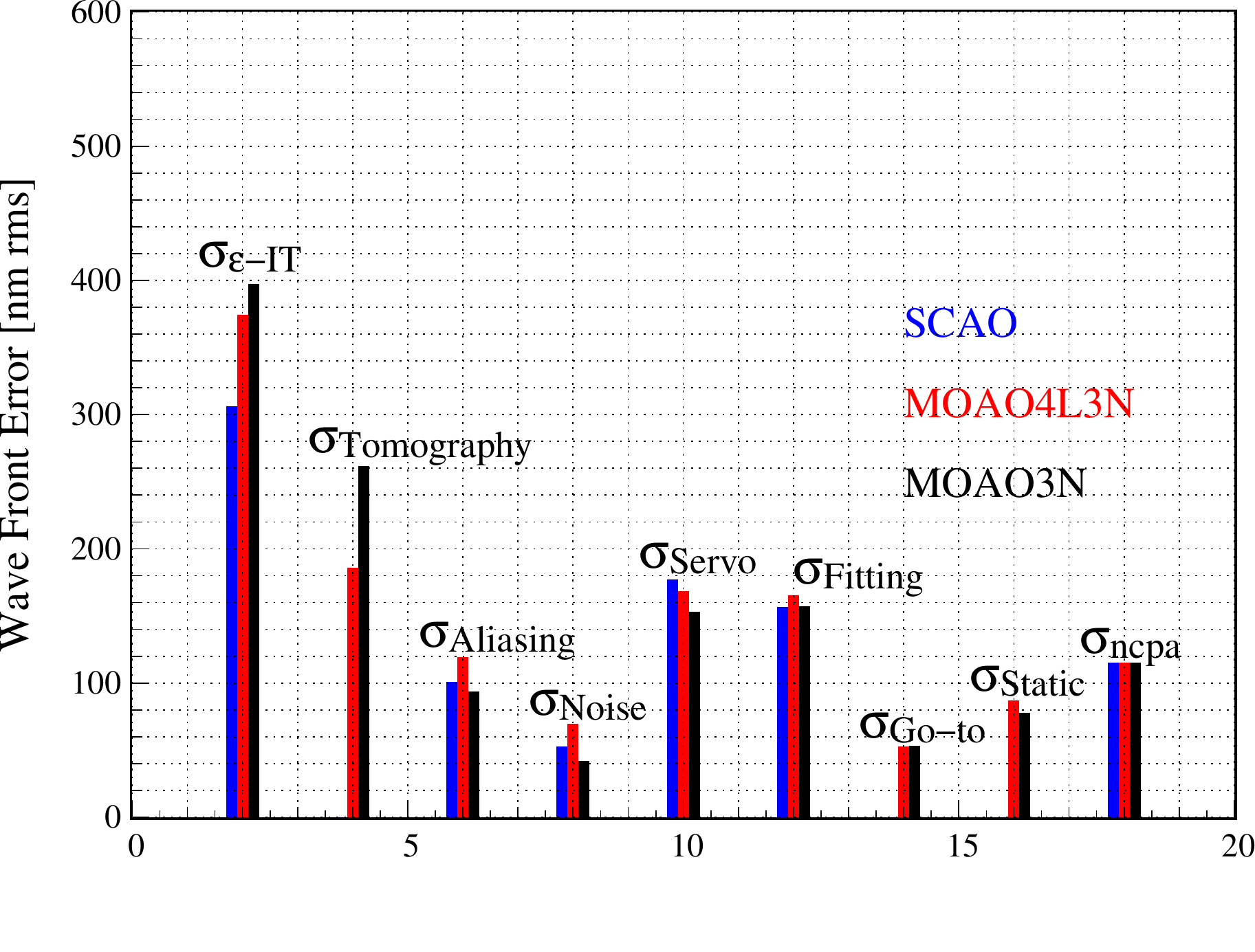}
		\caption{{Average error breakdown for each observation mode during script 275.}}
		\label{F:275_budget}
	\end{center}
\end{figure}

Script 275 was obtained directly after  script 274. The turbulent profile did not evolve significantly between the two scripts. We were comparing SCAO with MOAO, with and without LGS. We aim to evaluate the impact of using LGS on the tomographic reconstruction for the \cana case. 
We report in Fig.~\ref{F:275_budget} the error breakdown we get during this script. In MOAO 4L3N, we get the same errors compared to script 274 in Fig.~\ref{F:274_prof}, as we expect since the observational conditions were similar according to Table~\ref{T:scriptconditions} and the turbulent profile was similar as well.
In addition, we get very similar error values between MOAO 4L3N and MOAO 3N. The main difference comes from the tomographic error with 260~nm rms against 185~nm rms respectively in MOAO 3N and MOAO 4L3N.`
Overall, we see an increase of nearly 4\% in IR image SR~(see Table~\ref{T:scriptconditions}) when introducing LGS.

\begin{figure}
	\begin{center}
		\includegraphics[scale=.49]{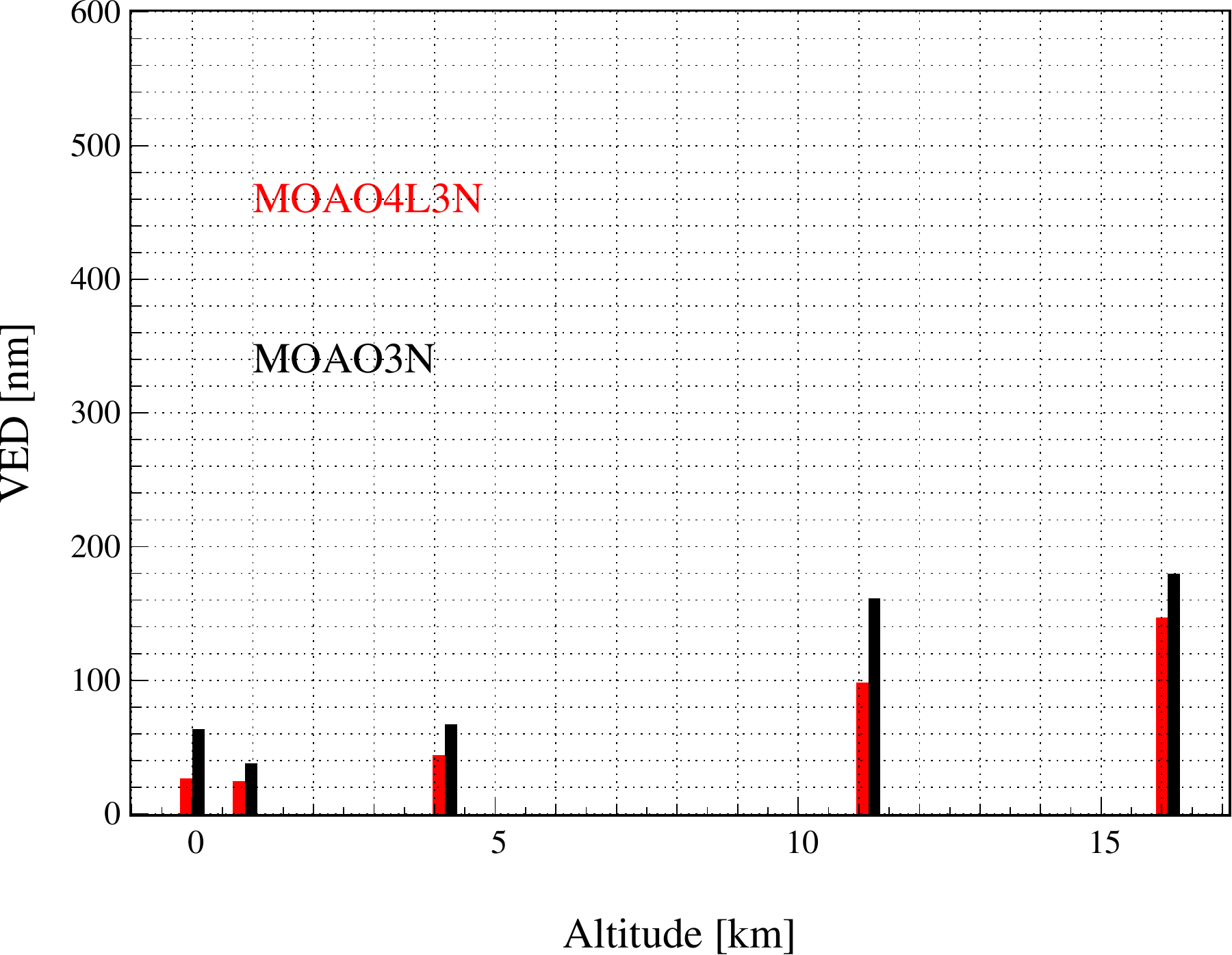}
		\caption{\small{Average VED~(nm rms) of the tomographic error along the altitude in MOAO during the script 275.}}
		\label{F:275_ved}
	\end{center}
\end{figure}

Fig.~\ref{F:275_ved} gives the VED of the tomographic error. Thanks to LGS, we reduce the error on every layer, even at the ground. The most substantial gain is obtained at 11km because of the relatively compact LGS asterism (LGS only 22.6" off-axis). While at 16 km, the low altitude of the LGS (21 km) limits the gain.  
The VED curve, as explained in~\cite{Gendron2014}, shows the distribution of the tomographic error along the altitude. In using more WFS, we should decrease the error level in this curve at any altitude. But, because of the cone effect, in adding LGS, the decrease of the tomographic error is not homogeneous over all altitudes, we so get a better relative improvement for low altitude layers than for high altitude ones. It is for this reason that we get such a gap for the 0 and 2~km layers between MOAO 4L3N and MOAO 3N. By using sodium LGS, we could expect to reduce this performance gap at higher altitudes.


\section{Conclusions}
\label{S6}

This paper gives a detailed analysis of \cana sky performance in its phase B configuration, for data sets acquired in 2013 at the WHT. In particular, we gave our method to manage the LGS tip-tilt filtering and how to compute mixed NGS plus LGS GLAO and MOAO reconstructors. We have showed the tomographic error can be computed from only a reconstructor and a covariance matrix of slopes, measured on uncompensated wave-fronts. This matrix can be produced either empirically or analytically~(see. appendix~\ref{A:1}). We have also proposed an analytic formulation of the residual phase variance. It is split into several error terms, which are assumed to be independent. We have detailed the calculation of each terms based on either the off-axis measurements or analytic formulas. We gave also another method to get the residual phase variance directly from the TS measurements. 
	
 Using 4,500 data sets, we gave statistics of \cana performance and presented average values of SR for different ranges of seeing, with 30.1, 21.4 and 17.1~\% of SR respectively in SCAO, MOAO and GLAO using 4 LGS and 3 NGS for median seeing. 
 
In addition, we have characterized the turbulence over the WHT during \cana on-sky observations. This was mostly dominated by the ground turbulence~(h<1~km), which reaches 0.59" of seeing~(500~nm) with a standard-deviation of 0.34", compared with 0.21" and 0.09" rms for combined higher altitude layers. The total turbulence has reached 0.66" of median seeing and 0.33" in standard-deviation. In addition, we have 0.09" of seeing above 20~km where \cana is not able to reconstruct the turbulence phase, but only for about 34\% of the time.
 
We have demonstrated that our analytic error breakdown computations follow the IR image measured SR fairly closely. Between the two error computation approaches (analytic and TS based), we get a correlation of 99~\%, for all AO modes. We have detailed the \cana performance on particular data sets. We have showed the LGS can increase the SR by up to 4\% compared to the using only the NGS, while the MOAO tomographic reconstructor allows a gain of 8.5 percentage points of SR, compared to GLAO compensation. 

We have also discussed the vertical error decomposition of the tomographic error, evaluating the gain brought by MOAO in altitude compared to GLAO, while the GLAO nulls the error in the ground layer. Moreover, we have highlighted the static aberrations in MOAO that must be carefully calibrated and monitored during the observations.

\section{Acknowledgements}
This work was supported by CNRS, INSU, Observatoire de Paris and  Universit\'e Paris Diderot-Paris 7 in France, Science and Technology Facilities Council and Univ. of Durham in UK, and European Commission (Fp7 Infrastructures 2012-1, OPTICON Grant 312430, WP1). These projects have also received the support of the A*MIDEX project (no. ANR-11-IDEX-0001-02) funded by the "Investissements d'Avenir" French Government programme, managed by the French National Research Agency (ANR). This work was also funded by the UK Science and Technology Facilities Council, grants ST/K003569/1 and ST/L00075X/1.  The raw data used in this publication is available from the authors.

\appendix
\section{Slopes covariance modelling}
\subsection{Parallel modes covariance}
\label{A:1}
We consider the $\xth{i}$ sub-aperture of the $\xth{p}$ WFS. In the pupil plane, we note $\ri$ the position of the $\xth{i}$ sub-aperture than does not depend on the WFS that are all conjugated at the pupil plane. Then, $\alphap$ will be the angular position of the $\xth{p}$ WFS from the center of the FoV. We will finally denote $\hlgsp$ the altitude focus of guide stars. We denote $\dipl$ the coordinate of this sub-aperture at the $\xth{l}$ layer in the Cartesian coordinate system of the pupil plane. We have~:
\begin{equation} \label{E:3}
\dipl= \left\lbrace
\begin{aligned}
& \alphap \hl + \ri\para{1 - \frac{ \hl }{\hlgsp}}\\
& \text{ for a LGS WFS and $\hl \leq \hlgsp$}.\\
& \alphap \hl + \ri \text{ else}
\end{aligned}
\right.
\end{equation}
For the following theoretical developments, we define $\xipl$ and $\yipl$ the projection of $\dipl$ on the pupil plane reference~:
\begin{equation} \label{E:4}
\dipl = \xipl \mathbf{x} + \yipl \mathbf{y},
\end{equation}
where $\mathbf{x}$ and $\mathbf{y}$ form an orthonormal basis along directions $x$ and $y$.
For a Shack-Hartmann WFS, the slopes 2D map comes from the average phase gradient over the lenslet~:
\begin{equation} \label{E:Sx0}
\mathbf{S}(\dipl) = \dfrac{1}{d_i}\iint_{\mathcal{A}} \boldsymbol{\nabla}\boldsymbol{\phi}(u_i,v_i) du_idv_i,
\end{equation}
where $d_i$ is the $\xth{i}$ sub-aperture pitch and $\mathcal{A}$ is the square integration domain comprised between $-d_i/2$ and $d_i/2$ in $x$ and $y$ directions. The integration variables $u_i$ $y_i$ are related to the $\xth{i}$ sub-aperture and are given by $u_i = x - \xipl$ and $v_i = y-\yipl$. 

We define the x-axis and y-axis slopes as the scalar product between $\mathbf{S}(\dipl)$ and respectively $\mathbf{x}$ and $\mathbf{y}$. In projecting the slopes along x or y directions, the integral on sub-aperture surface in Eq.~\ref{E:Sx0} becomes integrals along sub-aperture side~:
\begin{equation} \label{E:Sx}
\begin{aligned}
& \mathbf{S}(\dipl).\mathbf{x} = \dfrac{1}{d_i}\int\limits_{-d_i/2}^{d_i/2}  dv_i\para{\phi(\xipl+d_i/2,v_i) - \phi(\xipl-d_i/2,v_i)}\\
&\mathbf{S}(\dipl).\mathbf{y} = \dfrac{1}{d_i}\int\limits_{-d_i/2}^{d_i/2}  du_i\para{\phi(u_i,\yipl+d_i/2) - \phi(u_i,\yipl-d_i/2)}.
\end{aligned}
\end{equation}
We then define the spatial covariance $\gxx$, $\gyy$ and $\gxy$ as the following functions~:
\begin{equation}
	\begin{aligned}
	&\gxx(\Dijpql) = \aver{(\mathbf{S}(\dipl).\mathbf{x})\times (\mathbf{S}(\djql).\mathbf{x})  }\\
	&\gyy(\Dijpql) = \aver{(\mathbf{S}(\dipl).\mathbf{y})\times (\mathbf{S}(\djql).\mathbf{y})  }\\
	&\gxy(\Dijpql) = \aver{(\mathbf{S}(\dipl).\mathbf{x})\times (\mathbf{S}(\djql).\mathbf{y})  },
	\end{aligned}
\end{equation}
where $\Dijpql = \dipl - \djql$ is the separation between the $\xth{i}$ sub-aperture of $\xth{p}$ WFS and the $\xth{g}$ sub-aperture of $\xth{q}$ WFS, at altitude $\hl$. We develop here the method to get $\gxx(\Dijpql)$ only, both $\gyy(\Dijpql)$ and $\gxy(\Dijpql)$ are derived by the same way. The averaged gradient is the difference of phase along each side of the sub-aperture. We then get~:
\begin{equation} \label{E:SxSx}
\begin{aligned}
& \gxx(\Dijpql) = \dfrac{1}{d_i d_j} \iint dv_idv_j\\
& \left\langle \para{\phi(\xipl+d_i/2,v_i + \yipl) - \phi(\xipl-d_i/2,v_i+\yipl)}\right.\\
&\left.\para{\phi(\xjql+d_j/2,v_j+\yjql) - \phi(\xjql-d_j/2,v_j+\yjql)}\right\rangle. 
\end{aligned}
\end{equation}
The trick to handle Eq.~\ref{E:SxSx} is to use the following remarkable identity:
\begin{equation} \label{E:idenremar}
2\para{A -a}\para{B-b} =  -(A-B)^2 + (A-b)^2 +(a-B)^2 -(a-b)^2,
\end{equation}
that allows us to rewrite Eq.~\ref{E:SxSx} as follows~:
\begin{equation} \label{E:SxSx3}
\begin{aligned}
& \gxx(\Dijpql)= \frac{1}{2d_id_j} \iint dv_idv_j\\
& \left\langle -(\phi(\xipl+d_i/2,v_i+\yipl)-\phi(\xjql+d_j/2,v_j+\yjql))^2 \right.\\
& -(\phi(\xipl-d_i/2,v_i+\yipl)-\phi(\xjql-d_j/2,v_j+\yjql))^2\\
&+ (\phi(\xipl+d_i/2,v_i+\yipl) -\phi(\xjql-d_j/2,v_j+\yjql))^2\\
&\left. + (\phi(\xipl-d_i/2,v_i+\yjql) -\phi(\xjql+d_j/2,v_j+\yjql))^2\right\rangle\\
\end{aligned}
\end{equation}
Considering $\Dphi{\boldsymbol{\rho}} = \aver{(\phi(\mathbf{r})-\phi(\mathbf{r} + \boldsymbol{\rho}))^2}$ as the phase Structure Function~(SF), we have the final expression of the x-axis slopes covariance~: 
\begin{equation} \label{E:Gxx}
\begin{aligned}
& \gxx(\Dijpql) = \dfrac{1}{2d_id_j} \iint dv_idv_j\\
& \left(-2\Dphi{ \Dijpql + \dfrac{d_i-d_j}{2} \mathbf{x} + (v_i -v_j)\mathbf{y} }\right.\\
& +\Dphi{\Dijpql+ \dfrac{d_i+d_j}{2}\mathbf{x}+ (v_i -v_j)\mathbf{y} } \\
& \left.+ \Dphi{ \Dijpql - \dfrac{d_i+d_j}{2}\mathbf{x} + (v_i - v_j)\mathbf{y} } \right) . 
\end{aligned}
\end{equation}
We now consider $W_\phi(\kk)$ as the Von-K\'arm\'an spatial power spectral density~(PSD) given by~:
\begin{equation} \label{E:wphi}
W_\phi(\kk) = 0.023 \rz^{-5/3} \para{k^2 + 1/\lz^2}^{-11/6},
\end{equation}
The phase structure function~(SF) is related to the spatial PSD $W_\phi(\kk)$ by~:
\begin{equation} \label{E:dphi2}
\Dphi{\rhob} =  2\iint_{\mathbb{R}^2}W_\phi(\kk)\para{1 - cos(2\pi\kk\rhob)}d\kk.
\end{equation}
We commonly model the phase SF using the Von-K\'arm\'an expression~:
\begin{equation}\label{E:Dphi}
\begin{aligned}
& D_{\phi}(\boldsymbol{\rho}) = k_1 \para{\dfrac{\lz(\hl)}{\rz(\hl)}}^{5/3} \cro{k_2-  \rho^{5/6} K_{5/6}(\rho)},\\
& k_1 = \frac{2^{1/6} \Gamma(11/6)}{\pi^{8/3}} \cro{\frac{24}{5}\Gamma(6/5)}^{5/6}\\
& k_2 =  \frac{\Gamma(5/6)}{2^{1/6}},
\end{aligned}
\end{equation}
with $\rho = 2\pi \module{\Dijpql}/\lz(\hl)$. By the same way, we get~:
\begin{equation} \label{E:Gyy}
\begin{aligned}
&\gyy(\Dijpql)=\frac{1}{2d_id_j} \iint du_idu_j\\
& \left(-2\Dphi{ \Dijpql + \dfrac{d_i-d_j}{2}\mathbf{y} + (u_i -u_j)\mathbf{x}}\right.\\
& +\Dphi{\Dijpql+ \dfrac{d_i+d_j}{2}\mathbf{y}+ (u_i -u_j)\mathbf{x}}\\
&\left.+\Dphi{ \Dijpql - \dfrac{d_i+d_j}{2}\mathbf{y} + (u_i - u_j)\mathbf{x} } \right),
\end{aligned}
\end{equation}
and~:
\begin{equation} \label{E:Gxy} 
\begin{aligned}
&\gxy(\Dijpql)=\frac{1}{2d_id_j} \iint du_j dv_i\\
& \left(-\Dphi{ \Dijpql - \para{u_j-\dfrac{d_j}{2}}\mathbf{x} +  \para{v_i-\dfrac{d_i}{2}}\mathbf{y} }\right.\\ 
&- \Dphi{ \Dijpql -\para{u_j+\dfrac{d_j}{2}}\mathbf{x} + \para{v_i+\dfrac{d_i}{2}}\mathbf{y} } \\
&+\Dphi{ \Dijpql - \para{u_j-\dfrac{d_j}{2}}\mathbf{x} + \para{v_i+\dfrac{d_i}{2}}\mathbf{y}}\\
&\left.+\Dphi{ \Dijpql -\para{u_j+\dfrac{d_j}{2}}\mathbf{x} + \para{v_i-\dfrac{d_i}{2}}\mathbf{y}} \right) .
\end{aligned}
\end{equation}

Integrals in Eqs.~\ref{E:Gxx},~\ref{E:Gyy} and~\ref{E:Gxy} must be discretized along the sub-aperture side. To speed up the computation, we assume a Hudgin-like model~: the phase gradient is calculated in taking only the side-to-side phase difference at the middle of the sub-apertures. We have thus $u_i,u_j,v_i$ and $v_j$ forced to zero and the spatial covariance is obtained in summing the phase structure functions under integrals~(\cite{Gendron2014,Martin2012}).

\subsection{Aliasing covariance}
\label{A:2}

The covariance of the aliased phase can be derived using Eqs.~\ref{E:Gxx},~\ref{E:Gyy} and~\ref{E:Gxy}. These equations involve an expression of the phase SF, that depends on the turbulence characteristics and the geometry, according to Eq.~\ref{E:Dphi}. To get the aliasing contribution in the covariance, we have to split the phase into a parallel part, which is compensated by the system, and an orthogonal part~:
\begin{equation}
	\phi = \phi^\parallel + \phi^\perp.
\end{equation}
The orthogonal phase contains only spatial frequencies higher than $1/2d$, the DM cut-off frequency. 
Considering the quantity $\lz$ is large compared to $2d$, the aliasing SF $D^\perp_\phi(\rhob)$ is given by~:
\begin{equation} \label{E:dphialias}
D^\perp_\phi(\rhob) =  0.046 \rz^{-5/3}\iint \limits_{1/2d}^{\quad\infty} k^{-11/3} (1 -cos(2\pi \kk\rhob)) d\kk.
\end{equation}
Using Eq.~\ref{E:dphialias} to replace the SF expression in Eqs.~\ref{E:Gxx},~\ref{E:Gyy} and~\ref{E:Gxy}, we get an analytic expression of the spatial covariance of the aliasing.

\section{MOAO control}
\subsection{MOAO control law}
\label{B:1}

The residual modes in science directions are split into a compensated, or \emph{parallel} part by the DM, $\boldsymbol{\varepsilon_{\parallel}}$ and high frequencies non-reproducible, or \emph{orthogonal}, by the DM, $\boldsymbol{a_{\perp}}$~:
\begin{equation} \label{E:eps_bi}
\boldsymbol{\varepsilon} = \boldsymbol{\varepsilon_{\parallel}} + \boldsymbol{a_{\perp}}.
\end{equation}
The turbulence compensation achieved thanks to the DM results in that direction from the subtraction between the incoming parallel modes and the mirror modes $\boldsymbol{m_\text{DM}}$~:
\begin{equation} \label{E:eps_para_bi}
\boldsymbol{\varepsilon_{\parallel}} = \boldsymbol{a_{\parallel}} - \boldsymbol{m_\text{DM}}.
\end{equation}
The MOAO controller determines the mirror modes $\boldsymbol{m_\text{DM}}$ to apply at a sequence $t$ from the previous sequence $t-1$, the off-axis measurements $\sa$ and $\mathbf{M}_c$ the modal command matrix used on-sky. This later includes both tomographic reconstruction and DM/WFS calibrations~(\cite{Vidal2014}).
In order to get a system fractional delay $1+ \Delta_t$, we combine off-axis measurements acquired at $t-2$ and $t-1$ in this way~:
\begin{equation} \label{E:v_bi}
\begin{aligned}
&\boldsymbol{m_\text{DM}}(t) = (1-g)\times\boldsymbol{m_\text{DM}}(t-1) + \\
& g\mathbf{M}_c \mathbf{R} \para{\Delta_t\sa(t-2) + (1-\Delta_t)\sa(t-1) }.
\end{aligned}
\end{equation}
Applying the Z-transform to Eq.~\ref{E:v_bi}, we get~:
\begin{equation} \label{E:tmi}
\boldsymbol{\tilde{m}_\text{DM}}(z) = \tilde{h}_\text{ol}(z) \times\mathbf{M}_c \mathbf{R}\tilde{\s}_\text{off}(z)
\end{equation}
where $z = e^{-2i\pi\nu/ \nu_e}$ and $\nu_e$ the sampling frequency. The transfer function $\tilde{h}_\text{ol}(z)$ is the MOAO controller transfer function defined as~:
\begin{equation} \label{E:hol}
\tilde{h}_\text{ol}(z) =  g\times \dfrac{\Delta_t + (1-\Delta_t)z}{z(z-1+g)}
\end{equation}
The MOAO correction transfer function of the atmospheric parallel modes, $\tilde{h}_\text{cor}$ is given by~:
\begin{equation} \label{E:hcor}
\tilde{h}_\text{cor}(z) = \dfrac{\boldsymbol{\tilde{\varepsilon}_{\parallel}}}{\boldsymbol{\tilde{a}_{\parallel}} } = 1 - \tilde{h}_\text{ol}(z).
\end{equation}

\subsection{Disengaged TS measurements estimation}
\label{B:2}

We note $\sbe^\text{Eng}(t)$ the time series acquired by the TS in engaged loop. Using the voltages vector $\mathbf{V}$, the interaction matrix $\Mint$ and the calibrated the fractional delay $\tret$, the disengaged TS slopes is estimated by~:
\begin{equation} \label{E:sdis}
	\widehat{\s}_\text{on}(t) = \sbe^\text{Eng}(t) - \Mint\left(\tret\mathbf{V}(t-2) + (1-\tret)\mathbf{V}(t-1) \right).
\end{equation}
We reproduce the filtering operated by the RTC, described in Eq.~\ref{E:v_bi}, on-sky to estimate what the TS would measure if the DM stayed flat.

\end{document}